\documentclass[aps,prb,showpacs, amsmath,amssymb,floatfix,groupedaddress,twocolumn,superscriptaddress]{revtex4-2}
\usepackage{graphicx}
\usepackage{dcolumn}
\usepackage{bm}
\usepackage{times}
\usepackage[usenames,dvipsnames]{color}
\usepackage{colordvi,epsf}
\usepackage{multirow}
\usepackage{longtable}
\usepackage[utf8]{inputenc}
\usepackage[abs]{overpic}
\usepackage{colortab}

\usepackage{fixltx2e}
\usepackage{changes}
\usepackage{comment}

\usepackage{xspace}

\newcommand{\lsoc}{$\mathbf{L}^{\text{SOC}}$\xspace}
\newcommand{\ltom}{$\mathbf{L}^{\text{TOM}}$\xspace}
\newcommand{\ltot}{$\mathbf{L}^{\text{total}}$\xspace}

\newcommand{\ltotvert}{$\vert \mathbf{L}^{\text{total}} \vert$\xspace}
\newcommand{\lsocvert}{$\vert \mathbf{L}^{\text{SOC}} \vert$\xspace}

\usepackage{romannum}

\begin{document}
\pagenumbering{arabic}

\title{Topological and spin-orbit effects on orbital moments in ultra-thin magnetic films}
 
\author{Felix Nickel}
\email[Email: ]{nickel@physik.uni-kiel.de}
\affiliation{Institut f\"ur Theoretische Physik und Astrophysik,
Christian-Albrechts-Universit\"at zu Kiel, D-24098 Kiel, Germany}

\author{Soumyajyoti Haldar}
\affiliation{Institut f\"ur Theoretische Physik und Astrophysik,
Christian-Albrechts-Universit\"at zu Kiel, D-24098 Kiel, Germany}
\affiliation{Kiel Nano, Surface, and Interface Science (KiNSIS), University of Kiel, Germany}

\author{Mara Gutzeit}
\affiliation{Institut f\"ur Theoretische Physik und Astrophysik,
Christian-Albrechts-Universit\"at zu Kiel, D-24098 Kiel, Germany}

\author{Stefan Heinze}
\affiliation{Institut f\"ur Theoretische Physik und Astrophysik,
Christian-Albrechts-Universit\"at zu Kiel, D-24098 Kiel, Germany}
\affiliation{Kiel Nano, Surface, and Interface Science (KiNSIS), University of Kiel, Germany}

\date{\today}

\begin{abstract}

Topological orbital moments (TOMs) are a direct hallmark of a magnetic texture with a non-trivial spin topology. 
In addition to giving insight into the topology of the magnetic texture, TOMs could also be used to manipulate magnetic structures with a compensated total spin moment.
Experimental evidence of TOMs has been provided via
transport measurements of the Hall effect
in intercalated van-der-Waals materials.
However, a direct observation of TOMs is still missing.
Another complication arises for the unambiguous proof of the
topological origin since orbital moments can
also occur due to spin-orbit coupling.
Here, we use first-principles electronic structure theory to investigate the origin of orbital moments in 
different compensated spin structures with a non-trivial
topology.
We focus on ultrathin magnetic films at surfaces
such as Pd/Mn bilayers on Re(0001)
which represent ideal model systems for the detection of TOMs since it is possible to apply experimental 
techniques with local resolution of magnetic properties 
such as
spin-polarized scanning tunneling microscopy.

Due to its trivial topology we use the row-wise antiferromagnetic (RW-AFM) state
in a hexagonal monolayer
to analyze the spin-orbit induced contributions. 
The triple-Q (3Q) state is a superposition state of
three RW-AFM (1Q) states and thus electronically similar,
however, due to its non-trivial spin topology it
exhibits TOMs. 
The comparison between these two spin states allows us to disentangle the topological and spin-orbit contributions
to the orbital moments.
We find that TOMs have an important contribution in spin-compensated systems, since the spin-orbit coupling induced orbital moments are nearly compensated. First-principles calculations for atomic-scale 
skyrmion lattices in Fe monolayers on different surfaces
exhibit the same general trend found for the 3Q state.

\end{abstract}

\maketitle

\section{introduction}

Complex non-collinear magnetic structures have attracted considerable attention due to their rich physical properties and potential applications \cite{fert2013,Nagaosa2013,koraltan2026, rimmler2025}.
In recent years, there has been a particular focus 
on non-collinear spin structures with non-trivial topology on the atomic scale \cite{kurz2001,Taguchi2001,Martin2008, heinze2011, nakosai2013, Hoffmann2015, Hanke2016, nandy2016, DosSantosDias2016,Grytsiuk2020, Spethmann2020, gutzeit2023,nickel2025}. 
In many of these systems, the magnetic unit cell comprises only a few lattice sites, and the spin magnetization is fully compensated, meaning that the vector sum of all spin moments within the unit cell vanishes. Here, the spin moment refers to the magnetic moment arising from the spin momentum of the electrons.

In addition to spin moments, magnetic systems may also exhibit orbital magnetic moments originating from the orbital motion of electrons. These orbital contributions are often neglected because they are typically much smaller than the corresponding spin moments. However, in systems with compensated spin magnetization, orbital moments can become a decisive factor and may dominate the magnetic response of the material \cite{go2021}.

\begin{figure*}
    \centering
    \includegraphics[width=\linewidth]{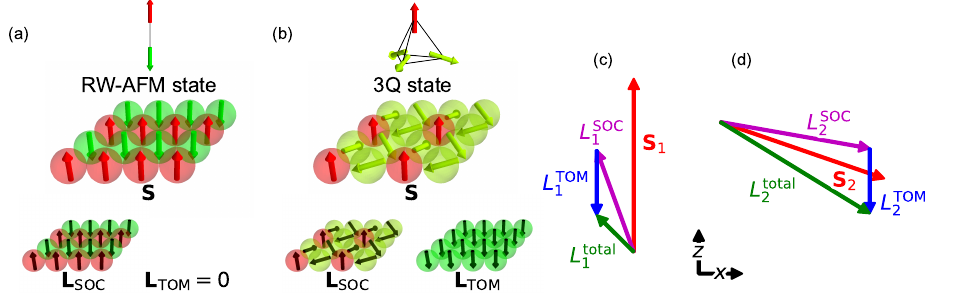}
    \caption{Schematics of the 
    contributions to the total orbital moment
    in an ultrathin film.
    (a) Spin structure of the RW-AFM state. The orbital moments due to SOC, \lsoc, can approximately be expected to be aligned with the spin moments $\mathbf{S}$. 
    Since the RW-AFM state is collinear the 
    topological orbital moments, \ltom, vanish.
    (b) Spin structure of the 3Q state with
    tetrahedron angles between adjacent spin
    moments in the 3Q$^1$ orientation with respect to
    the atomic lattice.  
    Besides \lsoc, which are nearly aligned with the spin moments, this state exhibits
    due to its non-trivial spin topology \ltom,
    which are uniform within the unit cell and
    aligned perpendicular to the film.
    (c+d) Spin moments, SOC induced orbital moments and TOM for two lattice sites of the 3Q$^1$ state. The SOC induced orbital moments and the TOMs add up to the total orbital moment. 
    The TOM is homogeneous over the spin structure, hence the same for both lattice sites (cf.~panel b). The magnitude of the SOC induced orbital moments and its angle to the spin moment depend on the absolute orientation of the spin moment. This leads to different contributions in (c) and (d).
    Orbital moments are shown schematically (not to scale) for visualization purposes.
    }
    \label{fig:intro}
\end{figure*}

In isolated atoms orbital moments arise due to the orbital quantum number of the occupied orbitals. This is a consequence of Hund's second rule to reduce the Coulomb interaction between the electrons. 
In solids, consisting of a very large number of atoms, this orbital moment is quenched due to the crystal field, as long as no global preferred direction exists.
Spin moments define such a preferred direction. Spin-orbit coupling (SOC) couples the orbital moments of all electrons to the magnetic moments, leading to a net orbital moment at each lattice site. 
Energy contributions arising from SOC can also be decisive for the magnetic ground state, e.g. due to the Dzyaloshinskii-Moriya interaction (DMI) or magnetocrystalline anisotropy energy (MAE). 
SOC-induced orbital moments occur in magnetic systems, independent of the topology of the magnetic configuration. 
As an example, we consider the spin structure of the
row-wise antiferromagnetic (RW-AFM) state in a hexagonal magnetic monolayer and its tentative orbital moments as sketched in Fig.~\ref{fig:intro}(a). 

Topological orbital moments (TOMs) 
represent a different type of orbital moments which have recently come into focus \cite{Hoffmann2015, Grytsiuk2020, Hanke2016, DosSantosDias2016, Hanke2017,Lux2018,Mankovsky2020,Mankovsky2021}. TOMs depend on the topology of the magnetic texture and occur only for a non-vanishing Berry-curvature. Therefore, the RW-AFM state exhibits no topological orbital
moments (Fig.~\ref{fig:intro}(a)).
A non-trivial topology arises for example for the non-coplanar 
triple-Q (3Q) state \cite{momoi1997, kurz2001, Spethmann2020, haldar2021, Nickel2023}, which exhibits tetrahedron angles between nearest-neighbor
spins as shown in Fig.~\ref{fig:intro}(b).
Here the TOM form an additional contribution, \ltom, besides the orbital moments due to SOC, \lsoc.
The total orbital moment is formed by the sum of the two terms
\begin{equation}
    \mathbf{L}^{\text{total}} = \mathbf{L}^{\text{SOC}} + \mathbf{L}^{\text{TOM}} \quad .
\end{equation}
Note that further contributions exist such as chiral orbital moments \cite{Lux2018}. However, it has been shown that these have only a small effect for a skyrmion lattice~\cite{Lux2018}. Therefore, further contributions are neglected in this work.
In an atomistic spin model the TOMs are given by \cite{Grytsiuk2020}
\begin{equation}
\label{eq:tom}
    \mathbf{L}_i^{\text{TOM}} = \sum_{(jk)} \kappa^{\text{TO}}_{ijk} \chi_{ijk} \pmb{\tau}_{ijk} \quad ,
\end{equation}
where $\kappa^{\text{TO}}_{ijk}$ is the topological orbital susceptibility, a material dependent parameter.
The non-coplanar alignment of the magnetic structure is described by the scalar spin chirality
\begin{equation}
\label{eq:ssc}
    \chi_{ijk} = \mathbf{s}_i \cdot (\mathbf{s}_j \times \mathbf{s}_k)  \quad,
\end{equation} 
defined for spin moments $\mathbf{s}_i$,  $\mathbf{s}_j$,
and  $\mathbf{s}_k$ at lattice sites $i$, $j$, and $k$.
For a collinear or coplanar spin texture the scalar
spin chirality is zero.
A non-zero scalar spin chirality results in a preferred direction for electrons to move around the three lattice sites \cite{Taguchi2001,tatara2003}.
This net chiral motion of electrons induces a topological orbital moment, which is perpendicular to the plane of the chiral current, i.e.~the plane of the three lattice sites. The direction of the TOM for each triplet is given by
\begin{equation}
    \pmb{\tau}_{ijk} \propto (\mathbf{r}_j-\mathbf{r}_i)\times(\mathbf{r}_k-\mathbf{r}_i) \quad
\end{equation}
with lattice vectors $\mathbf{r}_i$, $\mathbf{r}_j$,
and $\mathbf{r}_k$.
For a single magnetic layer the TOMs are consequently perpendicular to the surface, i.e.~either pointing up or down with respect to
the surface.
In contrast, \lsoc is usually nearly aligned with the magnetic moment.
For the non-coplanar 3Q state, the spin moments of the four atoms in the unit cell
compensate leading to a vanishing net spin moment. However, the TOMs are uniform
over the lattice resulting in a finite total orbital magnetization (Fig.~\ref{fig:intro}(b)). 

When the spin structure of the 3Q state is inverted, the scalar spin chirality, Eq.~(\ref{eq:ssc}), and consequently the TOMs,
Eq.~(\ref{eq:tom}), change sign and point into the opposite direction. Two domains, which are inverted with respect to each other have antiparallel aligned TOMs. 
These small orbital moments could be used to manipulate the domain by an external magnetic field even though the net spin moment vanishes. For example a magnetic field perpendicular to the surface
should shift the domain wall between two inverted domains in order to enlarge the domain, in which the TOMs are aligned with the field direction.

In the intercalated van-der-Waals material Co$_{1/3}$TaS$_2$ such a manipulation of the 3Q state has recently been reported by transport measurements of a
spontaneous topological Hall effect \cite{takagi2023, park2023}. 
Similar signatures in transport
experiments
have been found in the structurally similar material Co$_{1/3}$NbS$_2$ \cite{ghimire2018}, which were also related to the 3Q state \cite{khanh2025}. 
However, this provides only 
indirect evidence of TOM.
A net topological orbital magnetization is also predicted
for single skyrmions in a ferromagnetic background and it has
been suggested that these are observable via XMCD \cite{DosSantosDias2016}. 
However,
a direct measurement of TOMs has not been reported so far.

Ultrathin magnetic films on surfaces are good candidate systems to study TOMs locally, since they allow real-space observations, e.g.~via spin-polarized scanning tunneling microscopy (SP-STM) \cite{Spethmann2020,Nickel2023}. 
The magnetic state in such films can even be revealed
via SP-STM, when a monoatomic adlayer is grown on the magnetic layer
\cite{Romming2015,Meyer2019,Nickel2023}. Such an adlayer gives additional control over the 
electronic and magnetic
material properties of the system. 
However, as
discussed above,
TOMs do not occur in an isolated fashion but 
are accompanied with orbital moments originating from SOC.

Here we investigate the interplay between topological and spin-orbit coupling induced orbital moments in ultrathin magnetic films on surfaces based on density functional
theory (DFT). We consider the 3Q state in Pd/Mn/Re(0001) as an example for a 
non-coplanar spin structure with a non-trivial spin topology. The 3Q state in 
Pd/Mn/Re(0001) has been predicted based on DFT and experimentally observed via 
SP-STM \cite{Nickel2023}. Since this spin state can couple in different orientations to the atomic lattice it allows us to study different 
contributions from spin-orbit coupling to the orbital moments. We use the 
RW-AFM state as a reference state to disentangle orbital moments due to topology and spin-orbit coupling. Our analysis shows that the simple picture of additive
contributions to the total orbital moments holds very well for the 3Q state. 
This conclusion is also true for complex atomic-scale skyrmion lattice states
in Fe monolayers on the Re(0001) and Ir(111) surface covered by a few layers
of Ir and Rh, respectively.

This paper is structured as follows. In section \Romannum{2} we present the computational
details of our DFT calculations. 
In section \Romannum{3} we discuss the results of our
study. We begin with a comparison of the electronic structure of the collinear RW-AFM state and the non-coplanar 3Q state.
Next we analyze the SOC-induced orbital moments of the RW-AFM state 
and their dependence on the global orientation of the spin moments. 
The orbital moments of the electronically similar but non-coplanar 3Q spin state are discussed afterwards.
Based on these insights of the previous sections we present orbital moments for different nanoskyrmion lattice, where some are spin-compensated, while others are not. 
In section \Romannum{4} we provide the conclusions 
of our work.

\section{computational details}
We have performed DFT calculations using the \textsc{Fleur} code \cite{fleur-code, fleur-url},
which is based on the full-potential linearized
augmented plane wave (FLAPW) method \cite{wimmer1981, singh2005},
for a Pd/Mn/Re(0001) film system and a freestanding Pd/Mn/Re trilayer. 
The in-plane lattice constant is $2.76 \, \text{\AA}$, which is the experimental value for
Re. For both systems, 
DFT calculations were carried out in the RW-AFM and in the 3Q state.
The interlayer distances have been optimized for the Pd/Mn/Re(0001) film system in the RW-AFM state, in
which six layers of the Re substrate were considered. The three top layers are stacked in an A-B-C (fcc) stacking sequence, while the Re substrate has a C-A (hcp)
stacking sequence. Structural optimizations of the film system 
have been performed in the 
generalized gradient approximation (GGA) using the  PBE 
exchange-correlation functional \cite{Perdew1996}.
The relaxed interlayer distances obtained for
the Pd/Mn/Re(0001) film system are given in Ref.~\cite{Nickel2023}. The
interlayer distances of the freestanding 
Pd/Mn/Re trilayer were chosen according to
the relaxed values of the film system.
All DFT calculations apart from the structural relaxations were performed in 
the local density approximation (LDA) using the  exchange-correlation potential in the parameterization of Vosko, Wilk, and Nusair
\cite{vosko1980}.

The muffin-tin radii were chosen as $2.3 \, \text{a.u.}$ for Mn and Pd and $2.45 \, \text{a.u.}$ for Re. The cut-off parameter for the basis functions was 
$k_{\rm max}=4.1 \, \text{a.u.}^{-1}$. 
The RW-AFM state has been calculated in a $(2 \times 1)$ supercell, while the 3Q state was calculated in a $(2 \times 2)$ supercell. For both states a grid of $(25 \times 25)$ k-points has been used. 
All calculations, including those taking SOC into account, 
have been performed self-consistently.

All nanoskyrmion lattices were investigated with the \texttt{VASP} code \cite{vasp-url, kresse1996}, which is based on the projector augmented wave (PAW) method \cite{blochl1994,kresse1999}.  
The skyrmion lattice (SkX), the double-SkX (d-SkX), and the multi-Q state were calculated in a $16$ atom two-dimensional supercell for the film system
of an Fe monolayer (ML) on three atomic Ir layers on the Re(0001) surface, denoted
as Fe/Ir-3/Re(0001) in the following. A grid of $(15 \times 15 \times 1)$ 
k-points was used with an energy cut-off for the wave functions of 
$268 \, \text{eV}$. Further computational details can be found in Ref.~\cite{nickel2025}.
The SkX-12, SkX-19 and SkX-27 lattices were studied for an Fe ML
on the Ir(111) surface, an Fe ML on two atomic Rh layers on 
Ir(111), and for an Fe ML on one atomic Rh layer on Ir(111),
respectively. For these DFT calculations a cut-off of $300 \, \text{eV}$ has
been chosen and k-point meshes of $(10 \times 20 \times 1)$, $(11 \times 11 \times 1)$ and $(5 \times 15 \times 1)$, respectively, were used.
Further computational details can be found in Ref.~\cite{Gutzeit2022, gutzeit2023}.

\section{results}

First, we illustrate how topological and SOC contributions can combine to
the total orbital moment by a schematic example. We consider
a hexagonal magnetic monolayer
in the RW-AFM state (Fig.~\ref{fig:intro}(a))
and in the 3Q state (Fig.~\ref{fig:intro}(b)).
Fig.~\ref{fig:intro}(c,d) show two spin moments $\mathbf{S}_1$ and $\mathbf{S}_2$ from the 3Q$^1$ state (cf. Fig.~\ref{fig:intro}(b)).
$\mathbf{S}_1$ is aligned with the surface normal
($+z$-axis) and $\mathbf{S}_2$ has a tetrahedral angle with respect to it. 
Exemplary SOC induced orbital moments, \lsoc, and topological orbital moments, \ltom, are sketched for both orientations.
Since all neighbors in the 3Q state have the same angles between them and each triplet of neighbors has the same chirality, the TOM is homogeneous over the unit cell. This means all lattice sites have the same \ltom, in this case aligned along the $-z$-axis (
see lower right sketch in
Fig.\ref{fig:intro}(b)).
Note, that we assumed the sign of $\kappa^{\text{TO}}_{ijk}$ according to the Pd/Mn/Re(0001) film system (cf. Tab.~\ref{tab:tomSus}). For the opposite sign \ltom would be aligned along the $+z$-axis.
The SOC induced orbital moments, on the other hand, are typically roughly aligned with the spin moment. The magnitude of the SOC induced orbital moment and the angle with respect to the spin moment depend on the absolute direction of the spin moment. 
In Fig.~\ref{fig:intro}(c), the SOC induced orbital moment is chosen smaller than in Fig.~\ref{fig:intro}(d)
, since the magnitude of the orbital moment can in general depend on the orientation of the spin moment. 
$\mathbf{L}^{\text{SOC}}_1$ is nearly, but not perfectly aligned with $\mathbf{S}_1$ as is often the case. 
Note, that in the sketch
the size of the orbital moments is exaggerated for better visibility.
$\mathbf{L}^{\text{SOC}}_1$ and $\mathbf{L}^{\text{TOM}}_1$ are partially compensating, leading to a small total orbital moment. 
For $\mathbf{S}_2$ the SOC induced orbital moment, $\mathbf{L}^{\text{SOC}}_2$, is assumed to be larger (Fig.~\ref{fig:intro}d). Further, $\mathbf{L}^{\text{SOC}}_2$ and 
$\mathbf{L}^{\text{TOM}}_2$ both have negative $z$ components, 
resulting in a relatively large total orbital moment.
The two spin moments of the 3Q state demonstrate, that the SOC induced orbital moments are not homogeneous over the lattice and that the addition of \lsoc and \ltom can further enhance or decrease the total orbital moment at a given lattice
site.

Since \lsoc is nearly aligned with the spin moment at each atom site, one might expect the SOC induced orbital moments to cancel upon summing over the unit cell whenever the spin moments compensate. However, the magnitude of \lsoc varies between different sites as shown in Fig.~\ref{fig:intro}
and therefore, the net compensation is not guaranteed.
Further, the combination of \lsoc and \ltom raises the question, whether the TOM can be measured at all without explicit knowledge of the SOC induced orbital moments.
In the following, we address these issues based on electronic structure
theory via
DFT calculations for the example of the 3Q state of the freestanding Pd/Mn/Re trilayer and the Pd/Mn/Re(0001) 
film system and then generalize our conclusions to more complex nanoskyrmion spin structures in Fe-based 
ultrathin films.

\subsection{Electronic structure of the RW-AFM and 3Q state}
\begin{figure}
    \centering
    \includegraphics[width=\linewidth]{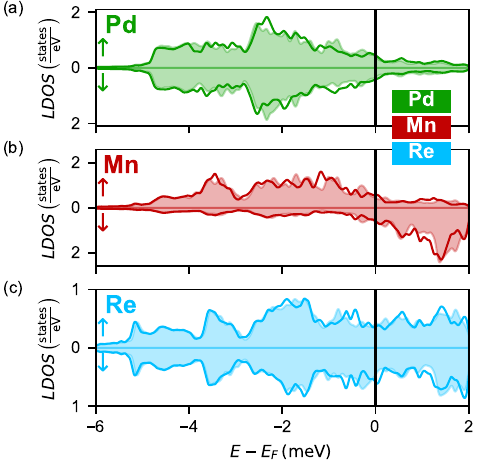}
    \caption{LDOS of a Pd/Mn/Re trilayer in the RW-AFM vs.~3Q state.
    The solid lines represent the LDOS of the freestanding
    Pd/Mn/Re trilayer
    obtained via DFT for the 3Q state 
    and the filled areas the LDOS of the RW-AFM state
    (cf.~Fig.~\ref{fig:intro}(a,b)). 
    The LDOS is shown for 
    (a) Pd, 
    (b) Mn and 
    (c) Re atoms. Note that these DFT calculations include SOC
    and the 3Q$^1$ orientation has been chosen for the 3Q state and the spin moments are oriented along the $z$-axis for the RW-AFM state. 
    While both atoms in the magnetic unit cell of the RW-AFM
    state have the same LDOS, the four lattice sites of the 3Q state differ slightly due to SOC. Here the lattice site with the spin moment pointing along the positive $z$-axis is shown, which is the same orientation shown for the RW-AFM state.
    }
    \label{fig:ldos}
\end{figure}

To investigate the contributions due to spin topology and spin-orbit coupling
to the orbital moments, we use the RW-AFM and 3Q state on a hexagonal lattice. These
spin states have a 2D magnetic unit cell of two and four atoms, respectively, making them well suited for a detailed analysis (cf.~Fig.~\ref{fig:intro}(a,b)). The RW-AFM state can be expressed as a spin spiral characterized by a $\mathbf{q}$-vector that is located at the $\overline{\text{M}}$ point of the 2D Brillouin zone. Therefore, this state can also be denoted as a single-Q (1Q) state. The 3Q state can be constructed as a superposition of 
three RW-AFM (1Q) states propagating along the three equivalent crystallographic
directions of the surface.
Both magnetic states have the same energy contribution due to exchange while the DMI vanishes \cite{haldar2021}. 
The 3Q state was found via DFT as the ground state in Pd/Mn/Re(0001) and experimentally
confirmed using SP-STM \cite{Nickel2023}. 

Here, as a model system for the
DFT calculations, we use in addition a freestanding Pd/Mn/Re trilayer, which allows for a detailed analysis of all three layers. Note, that we have
also performed DFT calculations for the
film system Pd/Fe/Re(0001) in which the Re
surface has been modeled by six Re layers 
(see Supplemental Material for details).

Fig.~\ref{fig:ldos} shows the local density of states (LDOS) for the Pd, Mn and Re atoms of the freestanding
Pd/Mn/Re trilayer. The solid lines represent the LDOS of the 3Q state, while the filled areas represent the RW-AFM state.
The shift between the LDOS of the spin-up and spin-down channels of Mn is the 
exchange splitting leading to a spin moment of 
$2.99 \, \mu_{\rm B}$.
For the Pd and Re atoms, there is also a small difference visible between the spin-up and spin-down channels, which corresponds to small spin moments 
of $0.07 \, \mu_{\rm B}$ and $0.03 \, \mu_{\rm B}$ per atom,
respectively, which are induced by the hybridization with the adjacent Mn atoms. 
For all three elements, the LDOS of the RW-AFM (1Q) state is quite similar to that
of the 3Q state consistent with their relation by the superposition.
Due to the close connection of their electronic structure, we expect that general trends of the orbital moments in the RW-AFM state are 
transferable to the 3Q state.
However, while the spin moments in the RW-AFM state are collinear, the 3Q state has a 
non-coplanar spin structure. Therefore, the RW-AFM state has only a spin-orbit
coupling contribution to the orbital moments, 
while the 3Q state exhibits both a spin-orbit and a topological contribution
to its orbital moments.

\subsection{Orbital moments of the RW-AFM state}
\begin{figure}
    \centering
    \includegraphics[width=\linewidth]{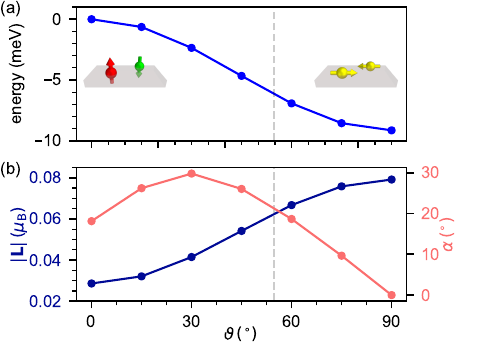}
    \caption{Total energy and orbital moments for the RW-AFM state. The spin moments in the RW-AFM state of a Pd/Mn/Re trilayer in fcc stacking
    are rotated from an out-of-plane orientation to 
    an in-plane orientation.
    (a) Total energy calculated via DFT including SOC. 
    (b) Magnitude of the orbital moments (blue) and angle $\alpha$ between the orbital and spin moments (red). 
    The dashed gray line indicates the angle, which corresponds to the 3Q$^2$ state (cf.~Fig.~5).
    }
    \label{fig:tri_rwafm_op_to_ip}
\end{figure}
To investigate the orbital contribution $\mathbf{L}^{\text{SOC}}$ independent of the topological contribution \ltom, we use the RW-AFM state
as a reference state. This collinear spin state has a trivial topology and therefore does not exhibit TOMs. In the RW-AFM state the spin moments form ferromagnetic rows, where neighboring rows are aligned antiparallel to each other, as shown in 
Fig.~\ref{fig:intro}(a). While keeping the angles between all spin moments fixed, the global spin orientation of the RW-AFM state can be varied.  
The optimal orientation of this state with respect to the atomic lattice is defined by effects of SOC, such as 
magnetocrystalline anisotropy energy
(MAE) and anisotropic symmetric exchange (ASE) \cite{Spethmann2020, haldar2021}.
Here we consider different orientations of the RW-AFM state, to observe how the total energy and the orbital moments depend on the orientation. 

In Fig.~\ref{fig:tri_rwafm_op_to_ip} the RW-AFM state is rotated from an out-of-plane ($\vartheta=0^{\circ}$) to an in-plane ($\vartheta=90^{\circ}$) orientation, 
i.e.~from an orientation of the spin moments perpendicular to the surface to an orientation parallel to the surface. The total energy obtained via
DFT (Fig.~\ref{fig:tri_rwafm_op_to_ip}(a)) decreases when the spin moments are rotated into the plane of the surface. Therefore, the MAE prefers an orientation
in the film plane,
a so called easy-plane anisotropy. As the energy decreases, the magnitude of the orbital moment increases (Fig.~\ref{fig:tri_rwafm_op_to_ip}(b)). This trend of a larger orbital moment for an energetically favored orientation due to the magnetocrystalline anisotropy is consistent with the well-known Bruno 
formula~\cite{Bruno1989}. 

The magnitude of the spin moment is nearly constant for all orientations.
While the magnitude of the orbital moments is increasing, the orbital moments are not always perfectly aligned with the spin moments. The angle $\alpha$ between the spin moments and the orbital moments is displayed in Fig.~\ref{fig:tri_rwafm_op_to_ip}(b). 
The angle $\alpha$ between $\mathbf{L}$ and $\mathbf{S}$ can be divided into a polar angle $\Delta \vartheta$ and an azimuthal angle $\Delta \varphi$.
As shown in Supplementary Fig. S1 and Supplementary Note 1, both $\Delta \vartheta$ and $\Delta \varphi$ are non-zero. 
For $\vartheta=90^{\circ}$, i.e.~an
in-plane orientation of the spin
moments,
$\mathbf{L}$ and $\mathbf{S}$ are perfectly aligned.

\begin{figure*}
    \centering
    \includegraphics[width=\linewidth]{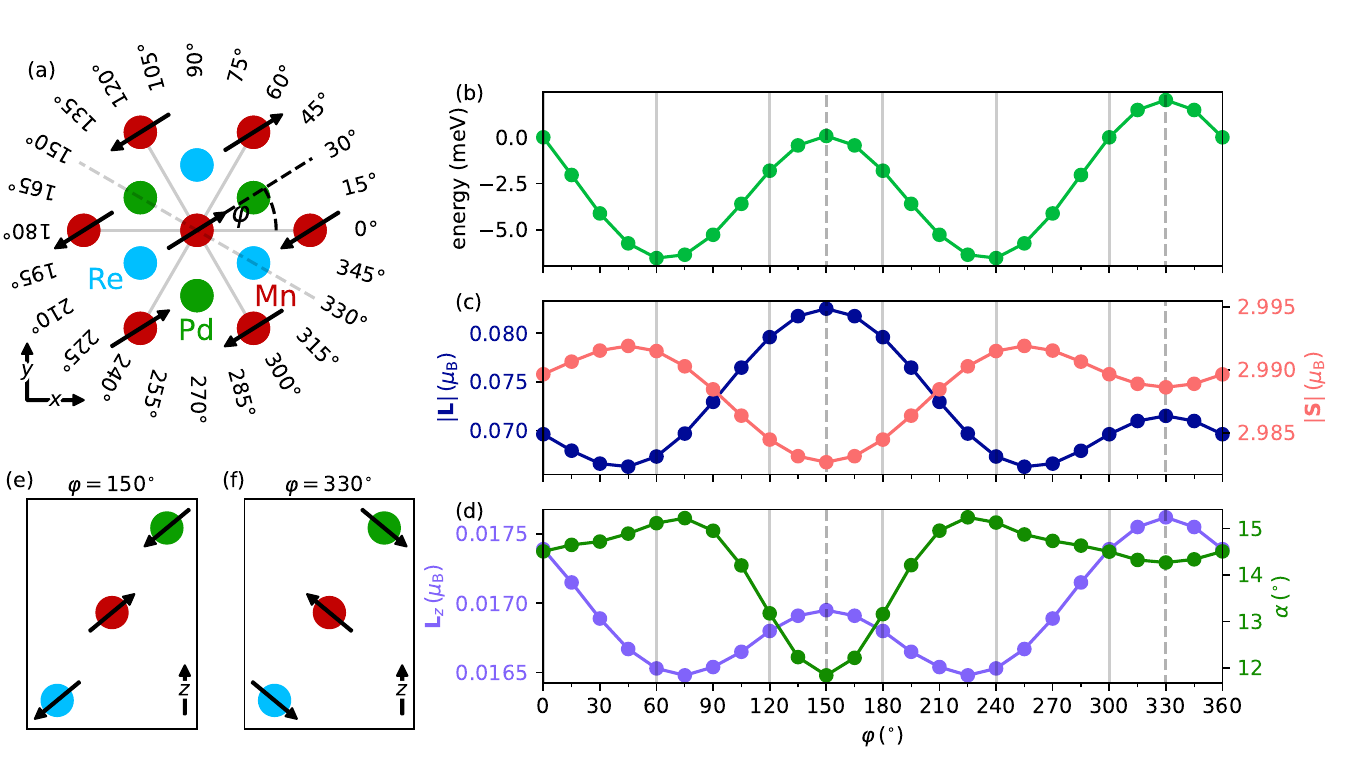}
    \caption{Orbital moments of the RW-AFM state for a varying azimuthal angle.
    The spin moments of the RW-AFM state in a Pd/Mn/Re trilayer have a polar angle of $\vartheta \approx 54.74^{\circ}$, which corresponds to the orientation of the spin moments fo the 3Q$^2$ state. This angle is indicated by a gray dashed line in Fig.~\ref{fig:tri_rwafm_op_to_ip}.
    The azimuthal angle $\varphi$ is varied.  
    (a) Sketch of the orientation of the spin moments with respect to the lattice for different angles $\varphi$. The gray dashed lines markes the angles represented in (e+f).
    (b) Total energy of system with respect to $\varphi=0$.
    (c) Magnitude of the orbital moment (blue) and the spin moment (red). 
    (d) $z$-component of the orbital moment (blue) and total angle $\alpha$ between the orbital and spin moment (green).
    (e+f) orientation of the spin moments for $\varphi = 150^{\circ}$ (e) and $\varphi = 330^{\circ}$ (f) for the three atoms, which are located on the dashed gray line in (a).
    }
    \label{fig:tri_rwafm}
\end{figure*}

In addition to the polar angle $\vartheta$, the azimuthal angle of the spin moments
$ \varphi$ can be varied. 
Upon changing $\varphi$, the MAE contribution can 
also vary with a three fold symmetry. 
For an unsupported magnetic monolayer the MAE would even have a six fold symmetry. 
However, the Pd and Re atoms
in the adjacent layers lower this symmetry to a 
three-fold rotational symmetry.
Nevertheless, this contribution is typically very small. A larger contribution can arise from the ASE, which 
also occurs due to SOC. In contrast to the MAE, the ASE is a two-site interaction.
In the atomistic spin model the ASE is given by 
\begin{equation}
    E_{\rm ASE} = - \sum_{i, j} J_{\rm ASE} (\mathbf{s}_i \cdot \mathbf{d}_{ij})(\mathbf{s}_j \cdot \mathbf{d}_{ij}) \quad ,
\end{equation}
where $\mathbf{d}_{ij}$ the normalized connection vector between lattice sites $i$ and $j$.
Depending on the sign of $J_{\rm ASE}$, which depends 
on the electronic structure of the system,
the ASE can favor an alignment of the RW-AFM state along or perpendicular to the ferromagnetic rows \cite{Spethmann2020}. For a positive sign, the ASE favors the same orientation as the 
magnetic dipole-dipole interaction.

In Fig.~\ref{fig:tri_rwafm} the spin moments of the RW-AFM state are oriented with a polar angle of $\vartheta \approx 54.74^{\circ}$ with respect to the surface
normal, which corresponds to the out-of-plane component of the 3Q$^2$ state (introduced below, cf.~Fig.~\ref{fig:tri_3q_lz}(c)). For comparison,
this angle is also indicated by a dashed gray line in Fig.~\ref{fig:tri_rwafm_op_to_ip}.
For this fixed polar angle the azimuthal angle $\varphi$ is varied, as shown in Fig.~\ref{fig:tri_rwafm}(a).
The total energy (Fig.~\ref{fig:tri_rwafm}(b)), including the effects of SOC, 
has two equivalent minima for $\varphi=60^{\circ}$ and $\varphi=240^{\circ}$.
The two minima indicate that the energy difference can be attributed to the ASE and not to the MAE, which would lead to at least three minima due to
symmetry (see Supplementary Note 2 for the MAE contribution).
In comparison to Fig.~\ref{fig:tri_rwafm_op_to_ip} the orbital moments (Fig.~\ref{fig:tri_rwafm}(c)) do not scale inverse proportional to the energy any more. This relation 
apparently only holds for the MAE and not for other SOC contributions, such as the two-site interaction ASE.
The angles of $\varphi=60^{\circ}$ and $\varphi=240^{\circ}$ correspond to 
an orientation of the spins parallel 
to the ferromagnetic rows of the RW-AFM state
(cf.~Fig.~\ref{fig:tri_rwafm}(a)),
which is the same ordering 
favored by the dipole-dipole interaction.
While the energy and the absolute value of the orbital moment do not scale inverse proportionally,
the absolute value of the orbital moment still varies with the direction of the spin moment.
\footnote{Note that when the spin moments
of the FM state, instead of the RW-AFM state, are rotated with $\varphi$, the two-site ASE has no contribution and also the variation of the orbital moment is negligible (see Supplementary Figure 2 and Supplementary Note 2).}.

For a rotation  of $\varphi$ by $180^{\circ}$ the orbital moment can differ. This can be seen for example for 
the angles of $\varphi=150^{\circ}$ and $\varphi=330^{\circ}$. For both angles the energy 
exhibits a local maximum, however, with a different
value (Fig.~\ref{fig:tri_rwafm}(b)).
The difference in energy comes along with a difference in the absolute value of the orbital moment as well as the spin moment ((Fig.~\ref{fig:tri_rwafm}(c)).
The origin of this energy difference for a rotation 
of $\varphi$ by $180^{\circ}$ lies in the orientation of the spin moments with respect to the position of the Re and Pd atoms. Figs.~\ref{fig:tri_rwafm}(e,f) show a side view of the atoms along the dashed gray line in Fig.~\ref{fig:tri_rwafm}a. For $\varphi=150^{\circ}$ the spin moments are roughly aligned with the connection vectors between the Mn atom and the Pd or the Re atom (Fig.~\ref{fig:tri_rwafm}(e)). As seen in Fig.~\ref{fig:tri_rwafm}(c), this leads to a larger orbital moment compared to the value for 
$\varphi=330^{\circ}$. Here, the spin moments are nearly perpendicular to the connection vectors (Fig.~\ref{fig:tri_rwafm}(f)).
This shows that both substrate and adlayer have an influence on the orbital moment of the Mn layer. 

While the absolute value of the orbital moments varies for a full rotation of $\varphi$ by about $0.016\, \mu_{\text{B}}$ (Fig.~\ref{fig:tri_rwafm}(c)), the variation with $\vartheta$ is significantly larger 
with a value of about $0.050\, \mu_{\text{B}}$ (Fig.~\ref{fig:tri_rwafm_op_to_ip}(b)).
The magnitude of the orbital moments varies up to $15 \%$ from the mean orbital moment. 
This demonstrates that the substrate and adlayer atoms do not have a large, but still a measurable effect on the orbital moments of Mn. Hence, the system can only be treated approximately as an effective monolayer system. 
Overall, the magnitude of the orbital moments depends on the alignment of the spin moments with each other as well as the alignment with respect to the induced spin moments of the Re and Pd atoms. 
The magnitude of the spin moments varies inverse proportional to the orbital moments.

The $z$-component of the orbital moment varies on the order of $0.001 \, \mu_{\text{B}}$ for a rotation
of the azimuthal angle (Fig.~\ref{fig:tri_rwafm}(d)).
There is also a difference between the values
for angles of $\varphi=150^{\circ}$ and 
$\varphi=330^{\circ}$. 
Therefore, the z-component does not only depend on the polar angle $\vartheta$, but also on the azimuthal angle $\varphi$. This is especially interesting with respect to the interplay of orbital moments due to SOC and due
to the spin topology studied for the 3Q state in the
next section.

\subsection{Orbital moments of the 3Q state}
The 3Q state is a superposition of three RW-AFM (1Q) states. As seen in Fig.~\ref{fig:ldos}, both have a similar electronic structure. Therefore, we expect that the general trends of the orbital moments due
to spin-orbit coupling, observed for the RW-AFM state, can be transferred to the 3Q state.
While the RW-AFM state is collinear, the 3Q state is non-coplanar and has therefore a non-trivial topology. 
In addition to the orbital moments due to SOC, this leads to the formation of TOMs, which are uniform over the whole unit cell (cf.~Fig.~\ref{fig:intro}(b)). It has been shown that the spin configuration of the 
RW-AFM (1Q) state can be continuously transformed into the 3Q state~\cite{haldar2021, nickel2025} and
that the TOMs are continuously increasing along the
path.

As seen for the RW-AFM state (Fig.~\ref{fig:tri_rwafm_op_to_ip}), the size of the orbital moment due to SOC depends on the orientation of the spin moment
with respect to the atomic lattice. Therefore, one expects the orbital moments to differ for individual lattice sites in
the unit cell of a non-collinear spin state.
For the 3Q state, both effects -- orbital moments due to SOC and due to the spin topology -- occur. 
To differentiate between them, one can calculate the TOM 
in DFT using the scalar-relativistic approximation, in which SOC effects are not included. 
An additional DFT calculation including SOC yields the total contribution of \lsoc and \ltom. When the TOM is already known, the SOC contribution to the orbital moments can be distinguished. 
However, the TOM depends on the electronic structure via the topological orbital susceptibility $\kappa^{\rm TO}$
(cf.~Eq.~(2)). In principle, a change in the electronic states, which occurs upon including SOC, can also modify 
the TOM. 

Typically, the influence of SOC on the electronic structure is 
small. This implies that the TOMs are relatively robust against the influence of small SOC effects. In Ref.~\cite{Lux2018} the TOMs were
explicitly calculated with gradually increasing SOC strength 
and found only small changes in the TOM for 
small SOC contributions.  
Here, we show that this conclusion also holds for several small-scale spin textures such as the 3Q 
state and nanoskyrmion lattices. 
\begin{figure*}
    \centering
    \includegraphics[width=\linewidth]{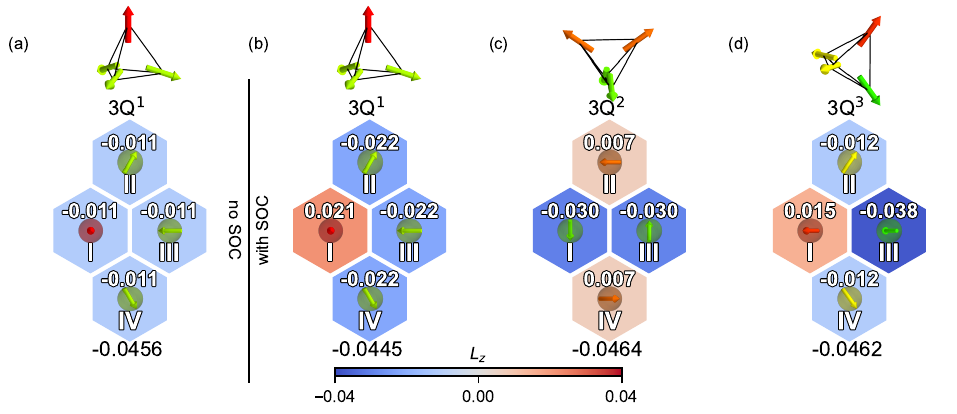}
    \caption{z-components of the total orbital moments for the 3Q state in different configurations.
    The z-component of the orbital moment for a Pd/Mn/Re trilayer,
    $L^{\rm total}_z$,
    is shown for three orientations of the 3Q state with respect to
    the atomic lattice: 
    (a+b) the 
    3Q$^1$, 
    (c) the 3Q$^2$, and 
    (d) the 3Q$^3$ orientation.
    The DFT calculations were performed for (a) in the scalar-relativistic approximation, i.e.~neglecting SOC, and for (b-d) including SOC. Note, that without SOC the orbital moments 
    originate only from the spin topology of the 3Q state
    and are invariant under a global spin rotation, i.e.~identical for the 3Q$^1$, 3Q$^2$ and 3Q$^3$ state. All values are given
    in units of $\mu_{\rm B}$.
    }
    \label{fig:tri_3q_lz}
\end{figure*}

In Fig.~\ref{fig:tri_3q_lz}(a) the orbital moments of the 3Q state are displayed for the 3Q$^{1}$ orientation with respect
to the lattice (indicated by the tetrahedron shown at the top
of the panel) at the different sites of the
unit cell by colored hexagons. In this calculation, SOC has
been neglected which results in a purely topological orbital
moment contribution.
At each lattice site the TOM has a magnitude of $0.011 \, \mu_{\text{B}}$
as obtained via our DFT calculations. A summation over the unit cell 
results in a total contribution of the TOM of $0.046 \, \mu_{\text{B}}$. 
These TOM only have a $z$-component and are invariant under global rotations. Therefore, all spin orientations of the 3Q state with respect to the lattice have the same TOM. 
While the TOM is invariant under a global spin rotation since
it depends on the spin topology, SOC-induced effects such as the ASE or \lsoc depend on the global orientation of the state. 

We consider three distinct spin orientations: the 3Q$^1$, the 3Q$^2$, and the 3Q$^3$ configuration (see Fig.~\ref{fig:tri_3q_lz}(b-d)). For the 3Q state, all neighboring spins span a tetrahedral angle. Therefore, this state can be visualized by a tetrahedron which can be used to indicate the orientation of the state relative to the atomic lattice (Fig.~\ref{fig:tri_3q_lz}).
The orientation of the TOM along the negative $z$-axis rather than the positive $z$-axis depends on the scalar spin chirality as well as the topological orbital susceptibility (cf.~Eq.~(\ref{eq:tom})). 
In a different material system
the TOMs for the same 3Q$^1$ state could be aligned with the $+z$-axis due to a change in the topological orbital susceptibility.
If, on the other hand, all spins are inverted, 
i.e.~going from an
all-out to an all-in 3Q state, the scalar spin chirality
changes sign (cf.~Eq.~(\ref{eq:ssc})) and 
all TOM point along the $+z$-axis.

Fig.~\ref{fig:tri_3q_lz}(b) shows the $L_z$ component of the orbital moment for the 3Q$^1$ orientation with SOC effects included. The total $L_z$ component summed over the unit cell is $0.045\, \mu_{\text{B}}$ and very similar to the TOM 
without SOC (Fig.~\ref{fig:tri_3q_lz}(a)).
Intuitively one can understand this, since the spin moments are compensated over the unit cell, i.e.~the total spin moment
vanishes, and the orbital moments \lsoc are roughly aligned with the spin moments, therefore also the \lsoc are compensated.
Interestingly this still holds true, if the magnitude of the orbital moments due to SOC varies over the unit cell, as our calculations show.
While the total $L_z$ contribution is very similar, the distribution within the unit cell is not any more uniform. 
When SOC is included the three lattice sites \Romannum{2} 
to \Romannum{4} have a negative contribution to the z-component of the
orbital moment, while the lattice site \Romannum{1} has a positive contribution. 

The absolute value of the orbital moment's z-component
for all four lattice sites is larger in the
3Q$^1$ orientation
upon including SOC (Fig.~\ref{fig:tri_3q_lz}(b)
than obtained neglecting it (Fig.~\ref{fig:tri_3q_lz}(a)). However, the contributions are partially compensating over the unit cell.
The variations at the different lattice sites originate from 
the orientation of the orbital moments due to SOC, \lsoc. 
In general, downward pointing spin moments have a larger $L_z$ component, since the TOM and orbital moment due to SOC point in a similar direction. This can be seen e.g.~for the atoms \Romannum{1} and \Romannum{3} of the 3Q$^2$ state (Fig.~\ref{fig:tri_3q_lz}(c)). Spin moments, which have a contribution along the positive $z$-direction, also have \lsoc with a positive $L_z$ component, which are partially
canceling
the TOM with a negative $z$-component, e.g.~atoms \Romannum{2} and \Romannum{4} of the 3Q$^2$ state (Fig.~\ref{fig:tri_3q_lz}(c)).
However, also for the 3Q$^2$ an 3Q$^3$ orientations we obtain a very similar total $L_z$ component upon summing over all atoms
in the unit cell (cf.~Fig.~\ref{fig:tri_3q_lz}(c,d)). Therefore,
we conclude from these calculations that the total orbital moment
of the unit cell of the 3Q state is nearly unaffected by
including the effect of SOC and remains basically at the
value obtained from the TOM.

\begin{figure*}
    \centering
    \includegraphics[width=0.9\linewidth]{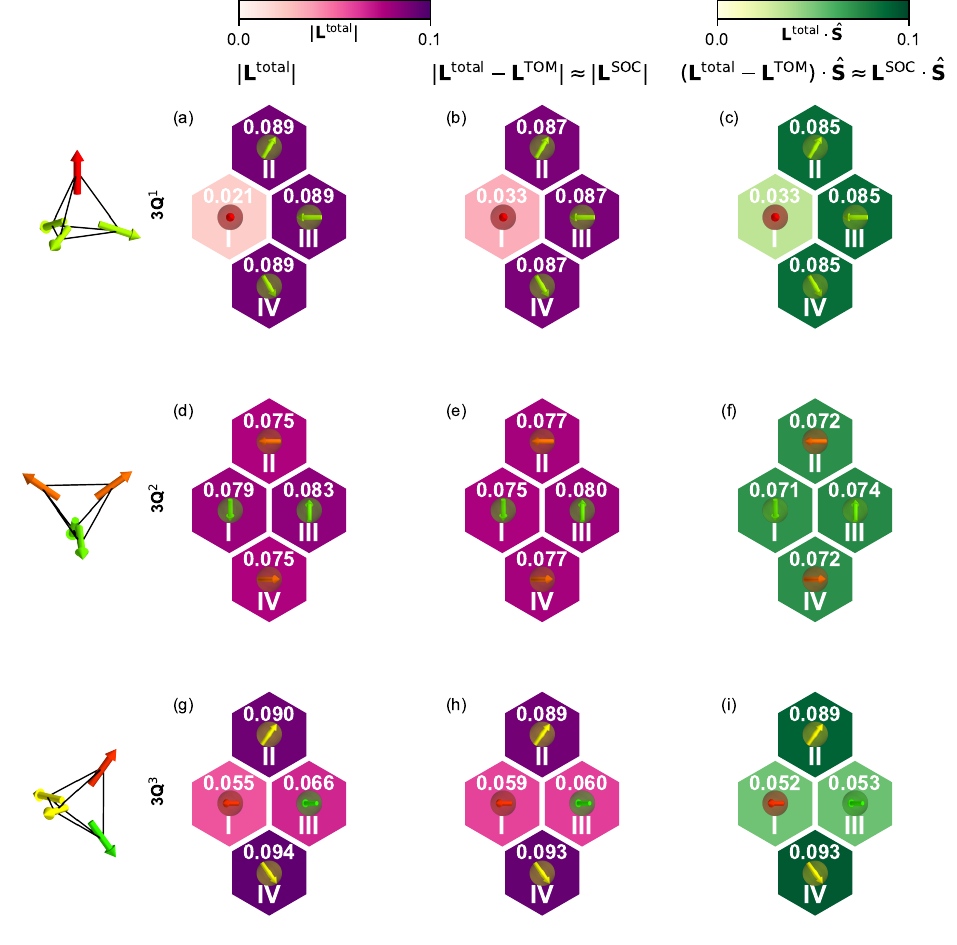}
    \caption{Orbital moments including SOC for three different orientations of the 3Q state.
    The orbital moments are shown for the Mn atoms of the
    freestanding Pd/Mn/Re trilayer.
    (a, d, g) Absolute value of the total orbital moment,
    $\mathbf{L}^{\rm total}$
    obtained via DFT, in which effects due to SOC and
    spin topology are included.
    (b, e, h) Absolute value of the orbital moment upon substracting the TOM obtained from the scalar-relativistic
    calculation (cf.~Fig.~\ref{fig:tri_3q_lz}a). This value is approximately the absolute value of \lsoc.
    (c, f, i) \ltom is subtracted from the total orbital moment, leaving approximately \lsoc. The resulting vector is projected onto the direction of the spin moment.
    All three quantities are displayed for the three considered orientations of the 3Q state with respect to the atomic
    lattice.
    (a-c) 3Q$^{1}$ state
    (d-f) 3Q$^{2}$ state
    (g-i) 3Q$^{3}$ state
    The \ltom contributions were obtained by a calculation without SOC (cf. Fig.~\ref{fig:tri_3q_lz}). All values are given
    in units of $\mu_{\rm B}$.} 
    \label{fig:tri_3q123_soc_tom}
\end{figure*}

To understand how spin-orbit coupling and spin topology contribute
to the total orbital moment, $\mathbf{L}^{\rm total}$,
we analyze the orbital moments for the 3Q$^1$, 3Q$^2$ and 3Q$^3$ configuration in more detail (Fig.~\ref{fig:tri_3q123_soc_tom}). 
Fig.~\ref{fig:tri_3q123_soc_tom}(a)  shows the absolute value of the total orbital moment of the Mn atoms in the Pd/Mn/Re trilayer
for the 3Q$^1$ state. For Mn atom \Romannum{1} the absolute value coincides with the $z$-component (cf. Fig.~\ref{fig:tri_3q_lz}(b)). Therefore, the orbital moment is completely aligned with the $z$-axis. As the total orbital moment for this lattice site is pointing along the positive $z$-axis, 
while the TOM points along the negative $z$-axis (Fig.~\ref{fig:tri_3q_lz}(a)), the contribution due to SOC needs to point along the positive $z$-axis and is stronger than the TOM contribution.
This alignment of \lsoc and \ltom is sketched in Fig.~\ref{fig:intro}(c). 
The alignment of \lsoc with the spin moment coincides with the results for the RW-AFM state (Fig.~\ref{fig:tri_rwafm}). This observation can be confirmed, when the TOM 
(Fig.~\ref{fig:tri_3q_lz}(a))
is subtracted from the total orbital moment, i.e.~by plotting $|\mathbf{L}^{\rm total}-\mathbf{L}^{\rm TOM}|$ (Fig.~\ref{fig:tri_3q123_soc_tom}(b)), 
which leaves approximately the SOC contributions. Note, that small differences 
could occur when the TOM obtained with and without SOC differ. 
Based on our DFT results we conclude that this effect is small.
As seen by comparing Fig.~\ref{fig:tri_3q_lz}(a) and Fig.~\ref{fig:tri_3q123_soc_tom}(b) the SOC contribution to the orbital moment of Mn atom \Romannum{1} is three times as large as the TOM contribution. 

In Fig.~\ref{fig:tri_3q123_soc_tom}(c), the difference between the total orbital moment and the TOM, $\mathbf{L}^{\rm total}-\mathbf{L}^{\rm TOM}$, is projected onto the local direction of the spin moment $\mathbf{\hat{S}}$. As discussed above, this is approximately the contribution of SOC to the orbital moment projected onto the direction of the spin moment, i.e.~$\mathbf{L}^{\rm SOC} \cdot \mathbf{\hat{S}}$. For Mn atom \Romannum{1} the projection and the absolute value of \lsoc are equal
(cf.~Fig.~\ref{fig:tri_3q123_soc_tom}(b) vs.~Fig.~\ref{fig:tri_3q123_soc_tom}(c)). Therefore, \lsoc is aligned with the spin moment and opposite to \ltom. Using the same quantities, we can also investigate the orbital moments of Mn atoms \Romannum{2} to \Romannum{4}. The absolute value of the total orbital moment (Fig.~\ref{fig:tri_3q123_soc_tom}(a)) is around four times larger than the $L_z$ component (Fig.~\ref{fig:tri_3q_lz}(b)). Thus in contrast to Mn
atom \Romannum{1}, the orbital moment is not aligned with the $z$-axis. This case is sketched in Fig.~\ref{fig:intro}d.
When the TOM is subtracted (Fig.~\ref{fig:tri_3q123_soc_tom}(b)), \lsocvert is only slightly smaller than $|\mathbf{L}^{\rm total}|$. 
Therefore, both contributions are not directly competing, as for Mn atom \Romannum{1}. 
When \lsocvert is compared to its projection onto the spin moment (Fig.~\ref{fig:tri_3q123_soc_tom}(c)) it becomes
apparent that \lsoc is nearly aligned with the spin moment. This is the expected orientation of the orbital moment due to SOC. 

In conclusion, the SOC contributions for Mn atom \Romannum{1} and atoms \Romannum{2} to \Romannum{4} differ, because they have a different out-of-plane component. This is consistent with the observations for the RW-AFM state (cf.~Fig.~\ref{fig:tri_rwafm_op_to_ip}).
Due to the opposite orientation of \lsoc and \ltom for Mn atom \Romannum{1}, both contributions are directly competing. As the spin moments of Mn atoms \Romannum{2} to \Romannum{4} are not directly aligned with the $z$-axis, the change of the magnitude of the orbital moments due to TOM is relatively small. As \lsoc, calculated from the difference between the total orbital moment and the TOM is consistent with the observations for the RW-AFM state, the approximation seems to hold that the TOM is nearly constant under the effect of SOC.

For the 3Q$^2$ configuration all Mn atoms have the same out-of-plane component (Fig.~\ref{fig:tri_3q123_soc_tom}(d)) . Therefore, all atoms are expected to have similar contribution to the orbital moments due to SOC. As seen from Fig.~\ref{fig:tri_rwafm} for the 
RW-AFM state, the \lsoc contribution can still differ slightly, depending on the azimuthal angle $\varphi$. From Fig.~\ref{fig:tri_3q123_soc_tom}(d) we find that indeed all atoms have a similar total orbital moment. The moments of Mn atoms \Romannum{2} and \Romannum{4} are exactly the same. The total
orbital moments for Mn atoms \Romannum{1} and \Romannum{3} are slightly larger 
and differ from each other. 
When the TOM contribution, obtained from the scalar-relativistic
DFT calculation (Fig.~\ref{fig:tri_3q_lz}(a)), is subtracted (Fig.~\ref{fig:tri_3q123_soc_tom}(e)), the orbital moment becomes slightly larger for Mn atoms \Romannum{2} and \Romannum{4}, while it is slightly decreased for atoms \Romannum{1} and \Romannum{3}. Since Mn atoms \Romannum{2} and \Romannum{4} have a positive $z$-component, the TOM reduces the total orbital moment. For Mn atoms \Romannum{1} and \Romannum{3}, with a negative $z$-component, the TOM increases the total orbital moment, as sketched in Fig.~\ref{fig:intro}(d). 
This can also be seen from Fig.~\ref{fig:tri_3q_lz}(c), where Mn atoms \Romannum{1} and \Romannum{3} have a $L_z$ component of $-0.030 \, \mu_{\text{B}}$. Here \lsoc and \ltom both point along the negative $z$-axis. For atoms \Romannum{2} and \Romannum{4}, \ltom is oriented along the negative $z$-axis and \lsoc exhibits a positive $z$-contribution such that the two contributions are competing, which results in $L_z = 0.007 \, \mu_{\text{B}}$.
The comparison between \lsocvert and the projection onto the direction of the spin moment (Fig.~\ref{fig:tri_3q123_soc_tom}(f)) demonstrates that all orbital moments are roughly, but not exactly aligned with the spin moments. This is expected, since for the RW-AFM state with the exact same out-of-plane component (Fig.~\ref{fig:tri_rwafm}(d)), the angle between orbital and spin moment varies between $12^{\circ}$ and $15^{\circ}$.

For the 3Q$^{3}$ configuration (Fig.~\ref{fig:tri_3q123_soc_tom}(g)) Mn
atoms \Romannum{2} and \Romannum{4} have a very similar magnitude of the orbital moment, while \ltotvert differs for atoms \Romannum{1} and \Romannum{3}. If the contribution of \ltom is subtracted (Fig.~\ref{fig:tri_3q123_soc_tom}(h)) the contributions of Mn atom \Romannum{1} and \Romannum{3} become nearly equivalent. This shows again that in one case the TOM is increasing the total orbital moment, 
while in the other case it is decreasing it.
The \lsocvert contribution for Mn atoms \Romannum{1} and \Romannum{3} is smaller than for Mn atoms \Romannum{4} and \Romannum{2}, as their
spin moments have an out-of-plane contribution, which leads to a smaller orbital moment. 
Due to their different spin alignment with the spin moments of the
Pd and Re atoms, Mn atom \Romannum{2} has a slightly smaller \lsoc than 
Mn atom \Romannum{4}.

As shown in Fig.~\ref{fig:tri_3q123_soc_tom}(i), the orbital moments of Mn atoms \Romannum{2} and \Romannum{4} align perfectly with their spin moments, while for 
Mn atoms \Romannum{1} and \Romannum{3} there is an angle between the orbital and the spin moment. This can be seen from the difference in Fig.~\ref{fig:tri_3q123_soc_tom}(h) and Fig.~\ref{fig:tri_3q123_soc_tom}(i). It is in line with the observations for the RW-AFM state since Fig.~\ref{fig:tri_rwafm_op_to_ip} shows that the orbital moments are nearly aligned with the spin moments, when the spin moment 
has no out-of-plane component.

Since the difference between \ltot and \ltom, calculated without SOC, is nearly aligned with the direction of the spin moments, one can conclude that $\mathbf{L}-\mathbf{L}^{\rm TOM} \approx \mathbf{L}^{\rm SOC}$.
However, there are small deviations, where the combination of TOM and SOC has a different contribution for two lattice sites, which would be symmetry equivalent under just one of the two effects. Examples for this are the lattice sites \Romannum{1} and \Romannum{3} of the 3Q$^{3}$ state.

Overall in the case of the 3Q state the $L_z$ component, summed over the unit cell, can be viewed as nearly identical without and with SOC considered. 
In both cases the unit cell has a net $z$-component of the orbital moment, which flips its sign when all spin moments are inverted. Potentially this allows to manipulate domains of the 3Q state by an external magnetic field which interacts with the orbital moments. Such an effect has been suggested by transport experiments performed for
intercalated van der Waals materials \cite{takagi2023,park2023, khanh2025}. Our DFT calculations indicate that the manipulation of the domains via a field 
is independent of SOC contributions to the orbital moments.  
On the other hand, the $L_z$ component of the orbital moments varies for different
atoms within the magnetic unit cell of the 3Q state, when SOC effects are considered. This could allow for a unique detection of the 3Q state and its orientation via STM. Harnessing the tunneling anisotropic magnetoresistance (TAMR) effect
\cite{Bode2002,Bergmann2012}, the 3Q state could be detected even with 
a non-magnetic STM tip.

In addition to the freestanding Pd/Mn/Re trilayer, which has been discussed so far, we have calculated the orbital moments for a Pd/Mn/Re(0001) film system. The orbital moments of the 3Q state with and without SOC are shown in Supplementary Figs.~S4 and S5. These figures are equivalent to Figs.~\ref{fig:tri_3q_lz} and \ref{fig:tri_3q123_soc_tom}, respectively, but consider the full film system, i.e.~take the effect
of the Re(0001) surface into account.
Qualitatively, the trilayer and the film system show the same behavior. The $z$ components of the total orbital moments, summed over the unit cell, are nearly identical with and without SOC (Supplementary Fig.~S4). 

Compared to the trilayer system, however, the TOMs are 
approximately twice as large.
For the film system, the ASE has the opposite sign compared to the 
freestanding trilayer, favoring an antiferromagnetic alignment along the connection vectors. Consequently, the angular dependence of the SOC-induced orbital moment differs between the two systems. Nevertheless, as in the trilayer, the total orbital moment in Pd/Mn/Re(0001) is approximately given by the sum of the SOC-induced orbital moment and the TOM calculated without SOC, further supporting our conclusions.

\subsection{Orbital moments of nanoskyrmion lattices}

\begin{figure*}
    \centering
    \includegraphics[width=\linewidth]{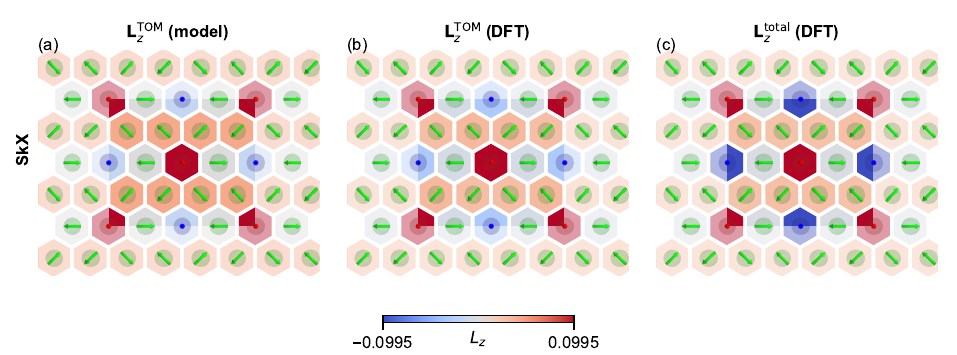}
    \caption{Topological and total orbital moments of the 
    SkX state in Fe/Ir-3/Re(0001).
    (a) Topological orbitals moments as calculated via Eq.~\eqref{eq:tom} and Eq.~(\ref{eq:ssc}), using the topological orbital susceptibility given in Tab.~\ref{tab:tomSus}.
    (b) $z$-component of the orbital moment obtained by a DFT calculation
    for Fe/Ir-3/Re(0001)
    neglecting SOC.
    The $x$- and $y$- components are zero for this calculation.
    (c) $z$-component of the orbital moment obtained by a DFT calculation including SOC. For this calculation the $x$- and $y$-components are locally non-zero.
    }
    \label{fig:skx_tom_soc}
\end{figure*}

The 3Q state is sometimes referred to as the smallest possible skyrmion lattice (SkX). We have shown previously that TOMs can also play a role in other atomic-scale skyrmion lattices \cite{nickel2025}. 
As an example,
an atomic scale SkX with zero net spin magnetization is displayed in Fig.~\ref{fig:skx_tom_soc}.
This non-coplanar spin state has been predicted as a possible ground state for an Fe monolayer on a Re(0001) surface covered
by three atomic layers of Ir -- denoted below as Fe/Ir-3/Re(0001) -- a system which has also been studied experimentally
via SP-STM \cite{nickel2025}.

Fig.~\ref{fig:skx_tom_soc}(a) shows the TOM calculated in the atomistic model for this spin texture, i.e.~based on Eq.~(\ref{eq:tom}).
In contrast to the 3Q state, the TOMs are not 
anymore uniform within the unit cell. For some lattice sites the TOM is pointing along the negative z-axis, for others the TOM points along the positive $z$-axis and for some the TOM is zero. 
Overall the lattice has a net TOM aligned with the positive $z$-axis.

\begin{table}[]
    \centering
    \begin{tabular}{c c c c c} \hline \hline
    system & state & $\kappa^{\rm TO}$ & $\frac{1}{N} \sum L^{\rm TOM}_z$  & $\frac{1}{N} \sum L^{\text{total}}_z$  \\ \hline
    Pd/Mn/Re(0001) & 3Q$^{1}$ & $ 0.005 $ &  $-0.021$ &  $-0.022$ \\
    Pd/Mn/Re(0001) & 3Q$^{2}$ & $0.005$  & $-0.021$  & $-0.020$ \\ 
    Pd/Mn/Re(0001) & 3Q$^{3}$ & $0.005$ & $-0.021$  & $-0.020$ \\ \hline
    Fe/Ir-3/Re(0001) & SkX & $-0.021$ & $0.022$ &  $0.021$ \\ 
    Fe/Ir-3/Re(0001) & multi-Q & $-0.017$ & $0.000$ &   $0.000$ \\
    Fe/Ir-3/Re(0001) & d-SkX & $-0.019$ & $0.000$ &  $0.000$ \\ \hline 
    Fe/Ir(111) & SkX-12 &  $ -0.021$ &  $0.042$ &  $0.019$ \\
    Fe/Rh-2/Ir(111) & SkX-19 &  $ -0.022$ &  $0.041$  & $0.023$\\
    Fe/Rh/Ir(111) & SkX-27 &   $ -0.023$ & $0.037$  & $0.021$ \\ \hline \hline
    \end{tabular}
    \caption{Topological orbital susceptibility and z-component of total orbital moments. The TOM have been calculated via
    DFT in the scalar-relativistic approximation for the Mn
    and Fe atoms, respectively, in the ultrathin film systems
    denoted in the 1st column for the magnetic state given
    in the 2nd column.  
    The topological orbital susceptibility $\kappa^{\rm TO}$ is assumed to be constant over the lattice and calculated by comparing the DFT calculated TOM with Eq.~\eqref{eq:tom} for the lattice site with the largest contribution.
    The TOM averaged over the $N$ atoms in the magnetic unit cell is given in the 4th column.
    The averaged z-component of the total orbital moments given
    in the last column
    has been calculated with the inclusion of SOC. 
    The spatial resolved $L^{\text{TOM}}_z$ and $L^{\text{total}}_z$ contributions are displayed in Fig.~\ref{fig:tri_3q_lz}, Fig.~\ref{fig:skx_tom_soc}, Supplementary Fig.~S6 and Supplementary Fig.~S7.
    All values are given in units of $\mu_{\rm B}$.}
    \label{tab:tomSus}
\end{table}

In general, according to Eq.~\eqref{eq:tom}, the topological orbital susceptibility can vary over the lattice. 
Here, we assume for simplicity
that the electronic structure is nearly constant 
over the lattice
and approximate the topological orbital susceptibility to be given
by a single constant $\kappa^{\rm TO}$
independent of the atom site.
We calculate the topological orbital susceptibility by comparing the TOM from DFT with the TOM from Eq.~\eqref{eq:tom}. Note, that the scalar-spin
chirality can be calculated based on the spin
structure alone (cf.~Eq.(\ref{eq:ssc})).
Therefore, we use the lattice site with the largest contribution to determine $\kappa^{\rm TO}$, 
in order to reduce numerical errors. The 
constant $\kappa^{\rm TO}$ obtained in this
way for Fe/Ir-3/Re(0001) is shown in Table \ref{tab:tomSus} and has been used to obtain
the TOM in the model ((Fig.~\ref{fig:skx_tom_soc}(a)). Note, that the value of $\kappa^{\rm TO}$
depends slightly on the considered spin state of the
Fe monolayer.

Fig.~\ref{fig:skx_tom_soc}(b) shows
the topological orbital moments for the 
atomic-scale SkX of Fe/Ir-3/Re(0001)
as calculated via DFT by neglecting SOC.
Since SOC has been excluded the orbital moments
can only have a topological origin.
The orbital moments are aligned along the
$z$-direction and nearly
coincide with the TOM, calculated by the atomistic model (Fig.~\ref{fig:skx_tom_soc}(a)). This demonstrates that the atomistic model gives a good description of the TOMs and
that the assumption of the spatially constant $\kappa^{\rm TO}$ is justified.
The $z$-component of the total orbital moment obtained
via DFT upon including SOC is shown in Fig.~\ref{fig:skx_tom_soc}(c). Fe/Ir-3/Re(0001) has an easy-axis magnetocrystalline anisotropy, meaning that an alignment parallel to the surface normal is favored \cite{nickel2025}. According to Ref.~\cite{Bruno1989} and Fig.~\ref{fig:tri_rwafm_op_to_ip} this should lead to larger orbital moments due to SOC for the magnetic moments, which are aligned with the surface normal. Indeed, we find that the magnetic moments pointing along $\pm z$ have the largest orbital moments (dark red and blue hexagons).

The $z$-component of the total orbital moments  (Fig.~\ref{fig:skx_tom_soc}(c))
are similar to the $z$-components
of the topological orbital moments (Fig.~\ref{fig:skx_tom_soc}(b)) but exhibit some slight differences, e.g. for the magnetic moments pointing along the negative
$z$-direction. The net $z$-component averaged over the whole unit cell (marked area) 
only changes from $0.022 \, \mu_B$ per atom without SOC to $0.021\,\mu_B$ per atom with SOC. 
In this case the net orbital moments are nearly unchanged, while locally the orbital moments differ when SOC is included or neglected. 
Tab.~\ref{tab:tomSus} displays the net orbital moments with and without SOC,
i.e.~total and topological orbital moments, for different film systems 
and magnetic states.
Here also the topological orbital susceptibilities are given. 
As seen from the three different states in the Fe/Ir-3/Re(0001) system, the topological orbital susceptibility can slightly differ between different magnetic states in the same system. In this case the variation is on the order of $10\%$.
The reason for this lies in the electronic structure, which changes slightly with the magnetic structure. 

Other non-coplanar spin states which have been 
studied for Fe/Ir-3/Re(0001)  
include a multi-Q and a double SkX state \cite{nickel2025}.
In these spin states 
the TOM exhibit an antiferromagnetic order and the net 
topological orbital magnetization vanishes (see 
Supplementary Fig.~7). Since the spin moments are compensated and the \lsoc are nearly aligned with the 
spin moments, the total orbital moments are also compensated.
The vanishing $L_z$ contribution of the total orbital moments
shows that also \lsoc is compensated, when the spin moment is compensated. 

For Fe monolayers on the clean and Rh covered Ir(111) surface
hexagonal skyrmion lattices have been previously studied
\cite{Gutzeit2022,gutzeit2023}. These SkX contain 12, 19,
and 27 atoms in the 2D unit cell and are therefore denoted
as SkX-12, SkX-19, and SkX-27 states (see Supplementary
Fig.~8 for sketches of the spin structures).
They are constructed in a similar way as the 3Q state 
by a superposition of three 1Q states. However,
in contrast to the 3Q state, here the period of the superimposed spin spirals is 
significantly larger ranging
from $8.1 \,$ nm to $12.2$ nm. 
Most importantly -- in contrast to the spin structures discussed so far --
these states are not spin compensated and therefore possess
a non-vanishing net spin moment. Consequently, also the orbital moment \lsoc is not compensated if we sum over all atoms in the unit cell. Therefore, in these 
skyrmion lattices, 
large differences occur in the net $L_z$ contribution of the topological
vs.~the total orbital moment (see table ~\ref{tab:tomSus}). 
In all spin compensated lattices, on the other hand, \lsoc is also nearly 
compensated and there is basically no difference between the averaged topological
vs.~total orbital moment (see first six lines in table ~\ref{tab:tomSus}). 
This makes the TOM in spin compensated systems still measurable, despite the 
effect SOC.

In general, Fe monolayer based material systems seem to have a roughly four times larger topological orbital susceptibility of 
$\approx -0.02 \, \mu_{\rm B}$ than the Mn based system Pd/Mn/Re(0001) with 
$\approx 0.005 \, \mu_{\rm B}$. Note also the opposite sign
between the two elements.

\section{Conclusion}
In this work, we have investigated via electronic structure theory based on DFT three different magnetic structures 
with a net zero spin moment: the RW-AFM state, the 3Q state, and the SkX state. Due to its collinearity the RW-AFM state exhibits no topological orbital moments,
while the 3Q and the SkX state are non-coplanar spin structures and possess homogeneous and non-homogeneous topological orbital moments, respectively. We have calculated via DFT the topological and spin-orbit coupling (SOC) induced contributions 
to the orbital moments in these spin structures.

From the calculations for the RW-AFM state, we conclude that \lsoc follows roughly the direction of the spin moments. The angle between both moments depends thereby on the direction of the spin moment and can be as large as $ \approx 30^{\circ}$. The norm of \lsoc scales with the polar angle $\vartheta$, but also with the azimuthal angle $\varphi$. The variation of \lsocvert with respect to $\varphi$ is thereby smaller than with respect to $\vartheta$. A $180^{\circ}$ rotation in $\varphi$ does not have to conserve the absolute value of \lsoc. The difference originates here from the alignment of the spin moment with the connection vectors to the substrate (Re) or adlayer (Pd).

For the 3Q state, we observe that the topological orbital moment calculated in the scalar-relativistic approximation, i.e.~neglecting 
SOC, 
is a very good approximation for the topological orbital moment obtained including SOC. Apparently, the SOC-induced changes of the electronic structure affect the topological orbital moment only in a negligible way. The \lsoc contribution differs for the atoms in the magnetic unit cell due to the different orientations of the spin moment. The dependence of the orbital moment on the spin direction is very similar to that observed for the RW-AFM state. Although the $L_z$ components for the individual lattice sites differ, the total $L_z$ component averaged over the unit cell is very similar, when only \ltom or when \ltom and \lsoc are considered.

Comparison between the atomistic spin model and DFT calculations for SkX states shows that $\kappa^{\text{TO}}$ is nearly constant over the lattice. This behavior is expected, since the electronic structure at the individual lattice sites varies only slightly. The comparison between different materials shows that Fe-monolayer based systems seem to have a larger $\kappa^{\text{TO}}$, than Mn-monolayer based systems.

Key findings of our work are that the magnitude of the SOC-induced orbital moment depends on the direction of the spin moment and that for spin compensated structures, the SOC induced orbital moments are also compensated over the unit cell. 
The magnitude of the TOMs is nearly not influenced by SOC effects.
For spin compensated magnetic structures, the topological orbital moment is
nearly identical to the
total orbital moment, when averaged over the unit cell. In this respect, 
spin-compensated states are much better suited for experimental studies
than spin-uncompensated states which exhibit in addition to the topological
part also a net orbital moment from SOC.
On the atomic scale, however, the total orbital moments differ from the 
topological orbital moments
by the SOC-induced orbital moments. This variation of the orbital moments
will also be reflected in the local electronic structure and may therefore
be detected in STM experiments.

It is our pleasure to thank Yuriy Mokrousov for insightful discussions.
We acknowledge financial support by the Deutsche
Forschungsgemeinschaft (DFG) via project no.~555842692 and computing time made
available to them on the high-performance computer "Lise" at the NHR
center NHR@ZIB. This center is jointly supported by the Federal Ministry
of Education and Research and the state governments participating in
the NHR (www.nhr-verein.de).

\bibliography{literature}

@article{vosko1980,
author = {Vosko, S. H. and Wilk, L. and Nusair, M.},
title = {Accurate spin-dependent electron liquid correlation energies for local spin density calculations: a critical analysis},
journal = {Can. J. Phys.},
volume = {58},
number = {8},
pages = {1200-1211},
year = {1980},
}

@article{Mankovsky2021,
  author    = {Mankovsky, Sergiy and Polesya, Svitlana and Ebert, Hubert},
  title     = {Topologically driven three-spin chiral exchange interactions treated from first principles},
  journal   = {Phys. Rev. B},
  volume    = {104},
  number    = {5},
  pages     = {054418},
  year      = {2021},
  month     = {Aug},
  doi       = {10.1103/PhysRevB.104.054418},
  publisher = {American Physical Society}
}

@article{Mankovsky2020,
  title = {Extension of the standard {H}eisenberg Hamiltonian to multispin exchange interactions},
  author = {Mankovsky, S. and Polesya, S. and Ebert, H.},
  journal = {Phys. Rev. B},
  volume = {101},
  issue = {17},
  pages = {174401},
  numpages = {21},
  year = {2020},
  month = {May},
  publisher = {American Physical Society},
  doi = {10.1103/PhysRevB.101.174401},
  url = {https://link.aps.org/doi/10.1103/PhysRevB.101.174401}
}

@article{Hanke2017,
  author    = {Hanke, Jan-Philipp and Freimuth, Frank and Bl{\"u}gel, Stefan and Mokrousov, Yuriy},
  title     = {Prototypical topological orbital ferromagnet $\gamma$-{FeMn}},
  journal   = {Sci. Rep.},
  volume    = {7},
  number    = {1},
  pages     = {41078},
  year      = {2017},
  month     = {Jan},
  doi       = {10.1038/srep41078},
  publisher = {Nature Publishing Group}
}

@article{wimmer1981,
  title = {Full-potential self-consistent linearized-augmented-plane-wave method for calculating the electronic structure of molecules and surfaces: ${\mathrm{O}}_{2}$ molecule},
  author = {Wimmer, E. and Krakauer, H. and Weinert, M. and Freeman, A. J.},
  journal = {Phys. Rev. B},
  volume = {24},
  issue = {2},
  pages = {864--875},
  numpages = {0},
  year = {1981},
  month = {Jul},
  publisher = {American Physical Society},
  doi = {10.1103/PhysRevB.24.864},
  url = {https://link.aps.org/doi/10.1103/PhysRevB.24.864}
}

@article{Bruno1989,
  title = {Tight-binding approach to the orbital magnetic moment and magnetocrystalline anisotropy of transition-metal monolayers},
  author = {Bruno, Patrick},
  journal = {Phys. Rev. B},
  volume = {39},
  issue = {1},
  pages = {865--868},
  numpages = {0},
  year = {1989},
  month = {Jan},
  publisher = {American Physical Society},
  doi = {10.1103/PhysRevB.39.865},
  url = {https://link.aps.org/doi/10.1103/PhysRevB.39.865}
}

@article{blochl1994,
  title = {Projector augmented-wave method},
  author = {Bl\"ochl, P. E.},
  journal = {Phys. Rev. B},
  volume = {50},
  issue = {24},
  pages = {17953--17979},
  numpages = {0},
  year = {1994},
  month = {Dec},
  publisher = {American Physical Society},
  doi = {10.1103/PhysRevB.50.17953},
  url = {https://link.aps.org/doi/10.1103/PhysRevB.50.17953}
}

@article{Perdew1996,
  title = {Generalized Gradient Approximation Made Simple},
  author = {Perdew, John P. and Burke, Kieron and Ernzerhof, Matthias},
  journal = {Phys. Rev. Lett.},
  volume = {77},
  issue = {18},
  pages = {3865--3868},
  numpages = {0},
  year = {1996},
  month = {Oct},
  publisher = {American Physical Society},
  doi = {10.1103/PhysRevLett.77.3865},
  url = {https://link.aps.org/doi/10.1103/PhysRevLett.77.3865}
}

@article{kresse1996,
  title = {Efficient iterative schemes for ab initio total-energy calculations using a plane-wave basis set},
  author = {Kresse, G. and Furthm\"uller, J.},
  journal = {Phys. Rev. B},
  volume = {54},
  issue = {16},
  pages = {11169--11186},
  numpages = {0},
  year = {1996},
  month = {Oct},
  publisher = {American Physical Society},
  doi = {10.1103/PhysRevB.54.11169},
  url = {https://link.aps.org/doi/10.1103/PhysRevB.54.11169}
}

@article{momoi1997,
  title = {Possible Chiral Phase Transition in Two-Dimensional Solid ${}^{3}\mathrm{He}$},
  author = {Momoi, Tsutomu and Kubo, Kenn and Niki, Koji},
  journal = {Phys. Rev. Lett.},
  volume = {79},
  issue = {11},
  pages = {2081--2084},
  numpages = {0},
  year = {1997},
  month = {Sep},
  publisher = {American Physical Society},
  doi = {10.1103/PhysRevLett.79.2081},
  url = {https://link.aps.org/doi/10.1103/PhysRevLett.79.2081}
}

@article{kresse1999,
  title = {From ultrasoft pseudopotentials to the projector augmented-wave method},
  author = {Kresse, G. and Joubert, D.},
  journal = {Phys. Rev. B},
  volume = {59},
  issue = {3},
  pages = {1758--1775},
  numpages = {0},
  year = {1999},
  month = {Jan},
  publisher = {American Physical Society},
  doi = {10.1103/PhysRevB.59.1758},
  url = {https://link.aps.org/doi/10.1103/PhysRevB.59.1758}
}

@article{kurz2001,
  title = {Three-Dimensional Spin Structure on a Two-Dimensional Lattice: {Mn/Cu}(111)},
  author = {Kurz, Ph. and Bihlmayer, G. and Hirai, K. and Bl\"ugel, S.},
  journal = {Phys. Rev. Lett.},
  volume = {86},
  issue = {6},
  pages = {1106--1109},
  numpages = {0},
  year = {2001},
  month = {Feb},
  publisher = {American Physical Society},
  doi = {10.1103/PhysRevLett.86.1106},
  url = {https://link.aps.org/doi/10.1103/PhysRevLett.86.1106}
}

@article{tatara2003,
  title = {Permanent current from noncommutative spin algebra},
  author = {Tatara, Gen and Kohno, Hiroshi},
  journal = {Phys. Rev. B},
  volume = {67},
  issue = {11},
  pages = {113316},
  numpages = {3},
  year = {2003},
  month = {Mar},
  publisher = {American Physical Society},
  doi = {10.1103/PhysRevB.67.113316},
  url = {https://link.aps.org/doi/10.1103/PhysRevB.67.113316}
}

@book{singh2005,
  author       = {Singh, David J and Nordstrom, Lars},
  title        = {Planewaves, Pseudopotentials and the LAPW Method, Second Edition},
  url          = {https://www.osti.gov/biblio/978055},
  place        = {United States},
  publisher    = {, Berlin, Germany},
  year         = {2005},
  month        = {01}}

@Article{heinze2011,
author={Heinze, Stefan
and von Bergmann, Kirsten
and Menzel, Matthias
and Brede, Jens
and Kubetzka, Andr{\'e}
and Wiesendanger, Roland
and Bihlmayer, Gustav
and Bl{\"u}gel, Stefan},
title={Spontaneous atomic-scale magnetic skyrmion lattice in two dimensions},
journal={Nat. Phys.},
year={2011},
month={Sep},
day={01},
volume={7},
number={9},
pages={713-718},
issn={1745-2481},
doi={10.1038/nphys2045},
url={https://doi.org/10.1038/nphys2045}
}

@article{nakosai2013,
  title = {Two-dimensional $p$-wave superconducting states with magnetic moments on a conventional $s$-wave superconductor},
  author = {Nakosai, Sho and Tanaka, Yukio and Nagaosa, Naoto},
  journal = {Phys. Rev. B},
  volume = {88},
  issue = {18},
  pages = {180503},
  numpages = {5},
  year = {2013},
  month = {Nov},
  publisher = {American Physical Society},
  doi = {10.1103/PhysRevB.88.180503},
  url = {https://link.aps.org/doi/10.1103/PhysRevB.88.180503}
}

@article{Hoffmann2015,
  title = {Topological orbital magnetization and emergent Hall effect of an atomic-scale spin lattice at a surface},
  author = {Hoffmann, M. and Weischenberg, J. and Dup\'e, B. and Freimuth, F. and Ferriani, P. and Mokrousov, Y. and Heinze, S.},
  journal = {Phys. Rev. B},
  volume = {92},
  issue = {2},
  pages = {020401(R)},
  numpages = {5},
  year = {2015},
  month = {Jul},
  publisher = {American Physical Society},
  doi = {10.1103/PhysRevB.92.020401},
  url = {https://link.aps.org/doi/10.1103/PhysRevB.92.020401}
}

@article{Hanke2016,
  title = {Role of {Berry} phase theory for describing orbital magnetism: From magnetic heterostructures to topological orbital ferromagnets},
  author = {Hanke, J.-P. and Freimuth, F. and Nandy, A. K. and Zhang, H. and Bl\"ugel, S. and Mokrousov, Y.},
  journal = {Phys. Rev. B},
  volume = {94},
  issue = {12},
  pages = {121114(R)},
  numpages = {5},
  year = {2016},
  month = {Sep},
  publisher = {American Physical Society},
  doi = {10.1103/PhysRevB.94.121114},
  url = {https://link.aps.org/doi/10.1103/PhysRevB.94.121114}
}

@article{nandy2016,
  title = {Interlayer Exchange Coupling: A General Scheme Turning Chiral Magnets into Magnetic Multilayers Carrying Atomic-Scale Skyrmions},
  author = {Nandy, Ashis Kumar and Kiselev, Nikolai S. and Bl\"ugel, Stefan},
  journal = {Phys. Rev. Lett.},
  volume = {116},
  issue = {17},
  pages = {177202},
  numpages = {5},
  year = {2016},
  month = {Apr},
  publisher = {American Physical Society},
  doi = {10.1103/PhysRevLett.116.177202},
  url = {https://link.aps.org/doi/10.1103/PhysRevLett.116.177202}
}

@Article{DosSantosDias2016,
author={dos Santos Dias, Manuel
and Bouaziz, Juba
and Bouhassoune, Mohammed
and Bl{\"u}gel, Stefan
and Lounis, Samir},
title={Chirality-driven orbital magnetic moments as a new probe for topological magnetic structures},
journal={Nat. Commun.},
year={2016},
month={Dec},
day={20},
volume={7},
number={1},
pages={13613},
issn={2041-1723},
doi={10.1038/ncomms13613},
url={https://doi.org/10.1038/ncomms13613}
}

@Article{ghimire2018,
author={Ghimire, Nirmal J.
and Botana, A. S.
and Jiang, J. S.
and Zhang, Junjie
and Chen, Y.-S.
and Mitchell, J. F.},
title={Large anomalous Hall effect in the chiral-lattice antiferromagnet {CoNb$_3$S$_6$}},
journal={Nat. Commun.},
year={2018},
month={Aug},
day={16},
volume={9},
number={1},
pages={3280},
issn={2041-1723},
doi={10.1038/s41467-018-05756-7},
url={https://doi.org/10.1038/s41467-018-05756-7}
}

@Article{Lux2018,
author={Lux, Fabian R.
and Freimuth, Frank
and Bl{\"u}gel, Stefan
and Mokrousov, Yuriy},
title={Engineering chiral and topological orbital magnetism of domain walls and skyrmions},
journal={Commun. Phys.},
year={2018},
month={Oct},
day={01},
volume={1},
number={1},
pages={60},
issn={2399-3650},
doi={10.1038/s42005-018-0055-y},
url={https://doi.org/10.1038/s42005-018-0055-y}
}

@article{Bode2002,
  title = {Magnetization-Direction-Dependent Local Electronic Structure Probed by Scanning Tunneling Spectroscopy},
  author = {Bode, M. and Heinze, S. and Kubetzka, A. and Pietzsch, O. and Nie, X. and Bihlmayer, G. and Bl\"ugel, S. and Wiesendanger, R.},
  journal = {Phys. Rev. Lett.},
  volume = {89},
  issue = {23},
  pages = {237205},
  numpages = {4},
  year = {2002},
  month = {Nov},
  publisher = {American Physical Society},
  doi = {10.1103/PhysRevLett.89.237205},
  url = {https://link.aps.org/doi/10.1103/PhysRevLett.89.237205}
}

@article{Meyer2019,
        title = {Isolated zero field sub{\textendash}{10} nm skyrmions in ultrathin {Co} films},
        author ={Meyer, Sebastian and Perini, Marco and von Malottki, Stephan and Kubetzka, Andre and Wiesendanger, Roland and von Bergmann, Kirsten and Heinze, Stefan},
        journal = {Nat. Commun.},
        volume = {10},
        number = {1},
        pages = {3823},
        year = {2019},
}

@article{Grytsiuk2020,
author = {Grytsiuk, S. and Hanke, J.-P. and Hoffmann, M. and Bouaziz, J. and Gomonay, O. and Bihlmayer, G. and Lounis, S. and Mokrousov, Y. and Bl{\"{u}}gel, S.},
title = {{Topological–chiral magnetic interactions driven by emergent orbital magnetism}},
year = {2020},
journal = {Nat. Commun.},
month = {dec},
number = {1},
pages = {511},
publisher = {Nature Research},
volume = {11},
doi = {10.1038/s41467-019-14030-3},
url = {http://www.nature.com/articles/s41467-019-14030-3}
}

@article{go2021,
doi = {10.1209/0295-5075/ac2653},
url = {https://doi.org/10.1209/0295-5075/ac2653},
year = {2021},
month = {sep},
publisher = {EDP Sciences, IOP Publishing and Società Italiana di Fisica},
volume = {135},
number = {3},
pages = {37001},
author = {Go, Dongwook and Jo, Daegeun and Lee, Hyun-Woo and Kläui, Mathias and Mokrousov, Yuriy},
title = {Orbitronics: Orbital currents in solids},
journal = {Europhys. Lett.}
}

@article{haldar2021,
  title = {Distorted $3Q$ state driven by topological-chiral magnetic interactions},
  author = {Haldar, Soumyajyoti and Meyer, Sebastian and Kubetzka, Andr\'e and Heinze, Stefan},
  journal = {Phys. Rev. B},
  volume = {104},
  issue = {18},
  pages = {L180404},
  numpages = {5},
  year = {2021},
  month = {Nov},
  publisher = {American Physical Society},
  doi = {10.1103/PhysRevB.104.L180404},
  url = {https://link.aps.org/doi/10.1103/PhysRevB.104.L180404}
}

@Article{takagi2023,
author={Takagi, H.
and Takagi, R.
and Minami, S.
and Nomoto, T.
and Ohishi, K.
and Suzuki, M.-T.
and Yanagi, Y.
and Hirayama, M.
and Khanh, N. D.
and Karube, K.
and Saito, H.
and Hashizume, D.
and Kiyanagi, R.
and Tokura, Y.
and Arita, R.
and Nakajima, T.
and Seki, S.},
title={Spontaneous topological {Hall} effect induced by non-coplanar antiferromagnetic order in intercalated van der {Waals} materials},
journal={Nat. Phys.},
year={2023},
month={Jul},
day={01},
volume={19},
number={7},
pages={961-968},
issn={1745-2481},
doi={10.1038/s41567-023-02017-3},
url={https://doi.org/10.1038/s41567-023-02017-3}
}

@Article{park2023,
author={Park, Pyeongjae
and Cho, Woonghee
and Kim, Chaebin
and An, Yeochan
and Kang, Yoon-Gu
and Avdeev, Maxim
and Sibille, Romain
and Iida, Kazuki
and Kajimoto, Ryoichi
and Lee, Ki Hoon
and Ju, Woori
and Cho, En-Jin
and Noh, Han-Jin
and Han, Myung Joon
and Zhang, Shang-Shun
and Batista, Cristian D.
and Park, Je-Geun},
title={Tetrahedral triple-Q magnetic ordering and large spontaneous {Hall} conductivity in the metallic triangular antiferromagnet {Co$_{1/3}$TaS$_2$}},
journal={Nat. Commun.},
year={2023},
month={Dec},
day={15},
volume={14},
number={1},
pages={8346},
issn={2041-1723},
doi={10.1038/s41467-023-43853-4},
url={https://doi.org/10.1038/s41467-023-43853-4}
}

@article{Taguchi2001,
  title = {Spin Chirality, {Berry} Phase, and Anomalous
{Hall} Effect in a Frustrated Ferromagnet},
  author = {Y. Taguchi and Y. Oohara and H. Yoshizawa and N. Nagaosa and Y. Tokura},
  journal = {Science},
  volume = {291},
  issue = {15},
  pages = {2573},
  numpages = {},
  year = {2001},
  month = {}
}

@article{Martin2008,
  title = {Itinerant Electron-Driven Chiral Magnetic Ordering and Spontaneous Quantum {Hall} Effect in Triangular Lattice Models},
  author = {Martin, Ivar and Batista, C. D.},
  journal = {Phys. Rev. Lett.},
  volume = {101},
  issue = {15},
  pages = {156402},
  numpages = {4},
  year = {2008},
  month = {Oct},
  publisher = {American Physical Society},
  doi = {10.1103/PhysRevLett.101.156402},
  url = {https://link.aps.org/doi/10.1103/PhysRevLett.101.156402}
}

@article{Nickel2023,
  title = {Coupling of the triple-$\mathrm{q}$ state to the atomic lattice by anisotropic symmetric exchange},
  author = {Nickel, Felix and Kubetzka, Andr\'e and Haldar, Soumyajyoti and Wiesendanger, Roland and Heinze, Stefan and von Bergmann, Kirsten},
  journal = {Phys. Rev. B},
  volume = {108},
  issue = {18},
  pages = {L180411},
  numpages = {6},
  year = {2023},
  month = {Nov},
  publisher = {American Physical Society},
  doi = {10.1103/PhysRevB.108.L180411},
  url = {https://link.aps.org/doi/10.1103/PhysRevB.108.L180411}
}

@Article{nickel2025,
author={Nickel, Felix
and Kubetzka, Andr{\'e}
and Gutzeit, Mara
and Wiesendanger, Roland
and von Bergmann, Kirsten
and Heinze, Stefan},
title={Antiferromagnetic order of topological orbital moments in atomic-scale skyrmion lattices},
journal={npj Spintronics},
year={2025},
month={Mar},
day={14},
volume={3},
number={1},
pages={7},
issn={2948-2119},
doi={10.1038/s44306-025-00074-3},
url={https://doi.org/10.1038/s44306-025-00074-3}
}

@Article{rimmler2025,
author={Rimmler, Berthold H.
and Pal, Banabir
and Parkin, Stuart S. P.},
title={Non-collinear antiferromagnetic spintronics},
journal={Nat. Rev. Mater.},
year={2025},
month={Feb},
day={01},
volume={10},
number={2},
pages={109-127},
issn={2058-8437},
doi={10.1038/s41578-024-00706-w},
url={https://doi.org/10.1038/s41578-024-00706-w}
}

@Article{khanh2025,
author={Khanh, Nguyen Duy
and Minami, Susumu
and Hirschmann, Moritz M.
and Nomoto, Takuya
and Jiang, Ming-Chun
and Yamada, Rinsuke
and Heinsdorf, Niclas
and Yamaguchi, Daiki
and Hayashi, Yudai
and Okamura, Yoshihiro
and Watanabe, Hikaru
and Guo, Guang-Yu
and Takahashi, Youtarou
and Seki, Shinichiro
and Taguchi, Yasujiro
and Tokura, Yoshinori
and Arita, Ryotaro
and Hirschberger, Max},
title={Gapped nodal planes and large topological Nernst effect in the chiral lattice antiferromagnet {CoNb$_3$S$_6$}},
journal={Nat. Commun.},
year={2025},
month={Mar},
day={26},
volume={16},
number={1},
pages={2654},
issn={2041-1723},
doi={10.1038/s41467-025-57320-9},
url={https://doi.org/10.1038/s41467-025-57320-9}
}

@misc{koraltan2026,
      title={The 2026 Skyrmionics Roadmap}, 
      author={Sabri Koraltan and Claas Abert and Manfred Albrecht and Maria Azhar and Christian Back and Hélène Béa and Max T. Birch and Stefan Blügel and Olivier Boulle and Felix Büttner and Ping Che and Vincent Cros and Emily Darwin and Louise Desplat and Claire Donnelly and Haifeng Du and Karin Everschor-Sitte and Amalio Fernández-Pacheco and Simone Finizio and Giovanni Finocchio and Markus Garst and Raphael Gruber and Dirk Grundler and Satoru Hayami and Thorsten Hesjedal and Axel Hoffmann and Aleš Hrabec and Hans Josef Hug and Hariom Jani and Jagannath Jena and Wanjun Jiang and Javier Junquera and Kosuke Karube and Lisa-Marie Kern and Joo-Von Kim and Mathias Kläui and Hidekazu Kurebayashi and Kai Litzius and Yizhou Liu and Martin Lonsky and Christopher H. Marrows and Jan Masell and Stefan Mathias and Yuriy Mokrousov and Stuart S. P. Parkin and Bastian Pfau and Paolo G. Radaelli and Florin Radu and Ramamoorthy Ramesh and Nicolas Reyren and Stanislas Rohart and Shinichiro Seki and Ivan I. Smalyukh and Sopheak Sorn and Daniel Steil and Dieter Suess and Mykola Tasinkevych and Yoshinori Tokura and Riccardo Tomasello and Victor Ukleev and Hyunsoo Yang and Fehmi Sami Yasin and Xiuzhen Yu and Chenhui Zhang and Shilei Zhang and Le Zhao and Sebastian Wintz},
      year={2026},
      eprint={2601.16575},
      archivePrefix={arXiv},
      primaryClass={cond-mat.mes-hall},
      url={https://arxiv.org/abs/2601.16575}, 
}

@misc{fleur-url,
    author = {},
  title = {{The FLEUR project}},
  howpublished = {\url{https://www.flapw.de/}},
}

@misc{fleur-code,
  author       = {Wortmann, Daniel and Michalicek, Gregor and Baadji, Nadjib and Betzinger, Markus and Bihlmayer, Gustav and Br\"oder, Jens and Burnus, Tobias and Enkovaara, Jussi and Freimuth, Frank and Friedrich, Christoph and Gerhorst, Christian-Roman and Granberg Cauchi, Sabastian and Grytsiuk, Uliana and Hanke, Andrea and Hanke, Jan-Philipp and Heide, Marcus and Heinze, Stefan and Hilgers, Robin and Janssen, Henning and Kl\"uppelberg, Daniel Aaaron and Kovacik, Roman and Kurz, Philipp and Lezaic, Marjana and Madsen, Georg K. H. and Mokrousov, Yuriy and Neukirchen, Alexander and Redies, Matthias and Rost, Stefan and Schlipf, Martin and Schindlmayr, Arno and Winkelmann, Miriam and Bl\"ugel, Stefan},
  title        = {{FLEUR}},
  month        = may,
  year         = 2023,
  publisher    = {Zenodo},
  doi          = {10.5281/zenodo.7576163},
  url          = {https://doi.org/10.5281/zenodo.7576163},
  howpublished  = {Zenodo}
}

@misc{vasp-url,
    author = {},
     title = {{VASP}},
	howpublished = {\url{https://www.vasp.at}},
}

@article{Gutzeit2022,
author = {Gutzeit, Mara and Kubetzka, Andr\'e and Haldar, Soumyajyoti and Pralow, Henning
and Goerzen, Moritz and Wiesendanger, Roland and Heinze, Stefan and von Bergmann, Kirsten},
year = {2022},
month = {},
journal ={Nat. Commun.},
volume = {13},
pages = {5764},
title = {Nano-scale collinear multi-Q states driven by higher-order interactions},
doi={https://doi.org/10.1038/s41467-022-33383-w}
}

@article{Nagaosa2013,
author = {Nagaosa, Naoto and Tokura, Yoshinori},
year = {2013},
month = {12},
pages = {899-911},
title = {{Topological properties and dynamics of magnetic skyrmions}},
volume = {8},
journal = {Nat. Nanotechnol.},
doi = {10.1038/nnano.2013.243}
}

@misc{FLEUR,
    author = {},
	address = {Forschungszentrum J{\"{u}}lich},
	howpublished = {See \url{https://www.flapw.de}},
}

@article{Spethmann2020,
author = {Spethmann, Jonas and Meyer, Sebastian and von Bergmann, Kirsten and Wiesendanger, Roland and Heinze, Stefan and Kubetzka, Andr\'e},
year = {2020},
month = {03},
pages = {227203},
title = {{Discovery of magnetic single- and triple-Q states in Mn/Re(0001)}},
journal= {Phys. Rev. Lett.},
 doi = {10.1103/PhysRevLett.124.227203},
volume = {124}
}

@article{Fert2013,
	author = {Fert, Albert and Cros, Vincent and Sampaio, Jo{\~{a}}o},
	isbn = {1748-3387},
	issn = {1748-3387},
	journal = {Nat. Nano.},
	month = {mar},
	number = {3},
	pages = {152--156},
	pmid = {23459548},
	title = {Skyrmions on the track},
	volume = {8},
	year = {2013}
}

@article{Romming2015,
	title = {{Field-Dependent Size and Shape of Single Magnetic Skyrmions}},
	author = {Romming, Niklas and Kubetzka, Andr\'e and Hanneken, Christian and von Bergmann, Kirsten and Wiesendanger, Roland},
	journal = {Phys. Rev. Lett.},
	volume = {114},
	issue = {17},
	pages = {177203},
	numpages = {5},
	year = {2015},
	month = {May},
	publisher = {American Physical Society},
}

@article{gutzeit2023,
  title = {Spontaneous square versus hexagonal nanoscale skyrmion lattices in {Fe/Ir}(111)},
  author = {Gutzeit, Mara and Drevelow, Tim and Goerzen, Moritz A. and Haldar, Soumyajyoti and Heinze, Stefan},
  journal = {Phys. Rev. B},
  volume = {108},
  issue = {6},
  pages = {L060405},
  numpages = {6},
  year = {2023},
  month = {Aug},
  publisher = {American Physical Society},
  doi = {10.1103/PhysRevB.108.L060405},
  url = {https://link.aps.org/doi/10.1103/PhysRevB.108.L060405}
}

@article{Bergmann2012,
  title = {Tunneling anisotropic magnetoresistance on the atomic scale},
  author = {von Bergmann, K. and Menzel, M. and Serrate, D. and Yoshida, Y. and Schr\"oder, S. and Ferriani, P. and Kubetzka, A. and Wiesendanger, R. and Heinze, S.},
  journal = {Phys. Rev. B},
  volume = {86},
  issue = {13},
  pages = {134422},
  numpages = {4},
  year = {2012},
  month = {Oct},
  publisher = {American Physical Society},
  doi = {10.1103/PhysRevB.86.134422},
  url = {https://link.aps.org/doi/10.1103/PhysRevB.86.134422}
}

\end{document}